\documentclass[9pt,twocolumn,twoside]{osajnl}

\journal{ao} 

\setboolean{shortarticle}{false} 

\usepackage{subcaption}
\usepackage{amsmath}

\title{SHADOWS: A spectro-gonio radiometer for bidirectional reflectance studies of dark meteorites and terrestrial analogues. Design, calibrations and performances on challenging surfaces.}

\author[1*]{Sandra Potin}
\author[1]{Olivier Brissaud}
\author[1,2]{Pierre Beck}
\author[1]{Bernard Schmitt}
\author[1]{Yves Magnard}
\author[1]{Jean-Jacques Correia}
\author[1]{Patrick Rabou}
\author[1]{Laurent Jocou}

\affil[1]{University Grenoble Alpes, CNRS, Institut de Planétologie et d'Astrophysique de Grenoble, 38000 Grenoble, France}
\affil[2]{Institut Universitaire de France, France}

\affil[*]{Corresponding author: sandra.potin@univ-grenoble-alpes.fr}

\ociscodes{(120.4570) Optical design of instruments, (120.5820) Scattering measurements, (240.649) Spectroscopy surface, (290.5880) Scattering, rough surfaces, (300.6340) Spectroscopy, infrared, (300.6550) Spectroscopy, visible}

 \doi{\url{https://doi.org/10.1364/AO.57.008279}}

\begin{abstract}
We have developed a new spectro-gonio radiometer, SHADOWS, to study in the laboratory the bidirectional reflectance distribution function of dark and precious samples. The instrument operates over a wide spectral range from the visible to the near-infrared (350-5000nm) and is installed in a cold room to operate at a temperature as low as -20°C. The high flux monochromatic beam is focused on the sample, resulting in an illumination spot of about 5.2mm in diameter. The reflected light is measured by two detectors with high sensitivity (down to 0.005$\%$ in reflectance) and absolute accuracy of 1$\%$. The illumination and observations angles, including azimuth, can be varied over wide ranges. This paper presents the scientific and technical constraints of the spectro-gonio radiometer, its design and additional capabilities, as well as the performances and limitations of the instrument.
\end{abstract}

\setboolean{displaycopyright}{true}

\begin{document}

\maketitle

\section{Introduction}

\subsection*{Scientific context}
\begin{figure}[h!]
\begin{center}
\includegraphics[scale=0.35]{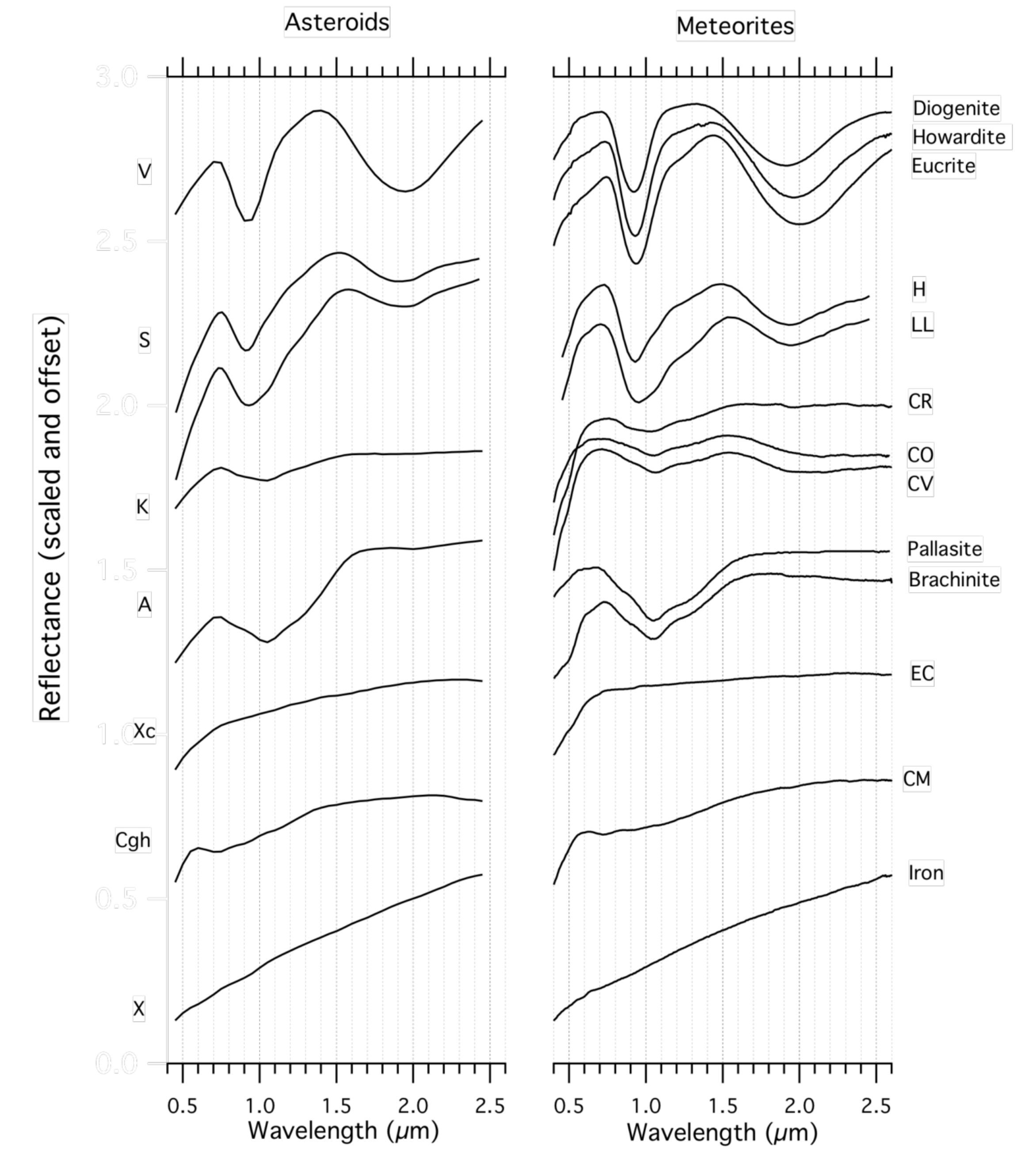}
\caption{Comparison between reflectance spectra of main belt asteroids and meteorites. Figure taken from \cite{small-bodies-pierres}.}
\label{compar asteroids météorites}
\end{center}
\end{figure}

Reflectance spectroscopy can provide information on the physical and chemical properties of surfaces of small bodies and planetary systems. This technique is currently used to classify asteroids according to the shape of their reflectance spectrum between 450 and 2450 nm, called the asteroid taxonomy\cite{bus,demeo_classification_asteroides}. Ground-based instruments and onboard space missions, such as New Horizons which crossed Pluto's system \cite{carte_pluton} or Rosetta who has orbited around the comet 67P/Churyumov-Gerasimenko for several years \cite{ciarniello}, have provided useful spectral data from asteroids, comets and planetary surfaces. As reflectance spectroscopy, especially with hyperspectral imaging, is a powerful investigative tool, this technique will continue to be widely used and much data is yet to come. The limits of this technique depend on the instrument spectral resolution, detector sensitivity, and on both surface illumination and albedo. Many solar-system surfaces, particularly primitive objects such as C- or D-type asteroids or comet nuclei, are extremely dark, presenting an albedo of a few percents in the visible, as shown by the spectra of the comet nucleus 67P/CHuryumov-Gerasimenko taken by the VIRTIS instrument onboard Rosetta \cite{composition-chury}. For all reflectance spectroscopy measurements, ground-based or onboard, the geometric configuration of the system is a major parameter to be taken into account for the analysis of the spectra. The reflectance of a surface generally depends on the geometry, thus presenting different spectral behaviors at different angles of illumination, observation or azimuth, as shown by numerous previous studies \cite{effet-opposition, H20_geometry, geometry}.\\
\hspace*{0.5cm}Laboratory measurements of meteorite samples or other planetary analogues are essential for simulating spectra acquired on the surfaces of small bodies, such as for matching asteroid spectra (see figure \ref{compar asteroids météorites}). In addition, the study of the bidirectional reflectance distribution function (BRDF) of meteorites can provide information on the surface texture, as has been demonstrated with the variations of the red slope of metallic meteorite spectra with different surface roughness \cite{brdf-metallic}. \\

Since these samples are quite precious and often exhibit less than 20$\%$ reflectance \cite{alex}, laboratory instruments must be capable of performing reflectance spectroscopy with a few milligrams of material, while accurately detecting very small light intensities. In addition, since the exact angular configuration may be essential to simulate a spectrum, these instruments must allow a wide range of angular configurations.\\

\subsection*{Objectives for SHADOWS}

SHADOWS, standing for \textit{Spectrophotometer with cHanging Angles for the Detection Of Weak Signals}, is especially designed to perform reflectance spectroscopy measurements of low albedo and precious samples, i.e. with small amounts of dark material. The performance target is to measure reflectances lower than 1$\%$ with less than a few mm$^3$ of material over most of the solar spectrum, 300 nm to 5 $\mu$m. We have also a particular interest in obtaining a high signal-to-noise ratio in the 3-4 $\mu$m range for the detection of organic and water related features. SHADOWS should be also fast enough to measure densely sampled BDRFs of dark meteorite samples and terrestrial analogues.\\
\hspace*{0.5cm}SHADOWS is based on our current spectro-gonio radiometer SHINE \cite{article_shine} but with a major difference in the illumination design. While SHINE sends a 200 mm wide collimated beam on the sample, SHADOWS focuses all the incident light into a spot of around 5 millimeters in diameter on the sample, thus considerably increasing the light flux density. Unlike SHINE where the detectors always see an homogeneously illuminated area, the SHADOWS illumination spot is always completely contained in the observation area. This reversed illumination-observation geometry should also allow accurate photometric measurements, but on small and low albedo samples. Like SHINE, SHADOWS offers a wide range of angular configurations, allowing an almost complete bidirectional coverage of reflectance measurements of the sample, and thus enabling bidirectional reflectance distribution function (BRDF) studies. The goniometer is installed in a cold room in order to acquire spectra at temperatures as low as -20°C. A high temperature static vacuum chamber and a cryogenic environmental cell integrated inside the goniometer are currently under development. Their goal is to offer a wide range of temperatures extending from -210°C to 400°C. The designs of these chambers are derived from SHINE's current static vacuum chamber and cryogenic environmental cell, respectively called Serac \cite{pommerol-martian-regolith} and CarboN-IR \cite{florence-carbonir}.\\
\hspace*{0.5cm}Finally, the control-acquisition software should be easy to use for visitors who are unfamiliar with reflectance measurements because the instrument is open as an European facility under the Europlanet program. The instrument must also be able to run several types of fully automated measurements to operate up to several days on long multi-angle spectral measurements (BRDF).

\subsection*{Existing instruments}
Several goniometers are already in use for specific applications. Categories emerge, such as the goniometers designed to study the opposition effect at low phase angle, opposed to goniometers with wider angular flexibility, or instruments to study the reflectance in situ opposed to laboratory instruments.\\
\hspace*{0.5cm} The Reflectance Eperiment LABoratory (RELAB) \cite{RELAB} currently located at the Johnson Space Center, Houston has an angular range from 0° to 60° for the incidence and measurement angles. The unpolarized monochromatic light is generated by a monochromator and two cooled detectors, one for the visible and one for the infrared range, measure the reflected intensity. The nominal configuration of this goniometer allows measurements with a spectral sampling of 1 nm and a spectral resolution around 1 nm from 400 to 2700 nm.\\
\hspace*{0.5cm} The European Goniometric Facility (EGO) \cite{EGF} at the Institute for Remote Sensing Applications of the Joint Research Centre at Ispra, Italy consists of two horizontal circular rails and two vertical arcs. Each arc is motorized and enables the positionning of the light source and the detector anywhere on a 2m radius hemisphere. The light source and detectors can be changed according to the experiment needs.\\
\hspace*{0.5cm} The FIeld Goniometer System FIGOS \cite{FIGOS} is a transportable system made for in-situ measurements. With one azimuth circular rail and one zenith arc, both with a 2m radius, the goniometer allows a full 360° rotation, and can acquire reflectance measurements from 300 to 2450 nm.\\
\hspace*{0.5cm} The long-arm goniometer of the Jet Propulsion Laboratory in Pasadena \cite{long-arm_goniometer} is especially designed to measure the opposition effect, with a minimum phase angle of 0.05°. The light source is a HeNe laser at 632 nm modulated by a chopper and a p-i-n diode catches the reflected light at the end of a movable arm. Two quarter wave plates can be placed in the optical path to analyse the circular polarization of the reflected light.\\
\hspace*{0.5cm} The SpectropHotometer with variable INcidence and Emergence SHINE \citep{article_shine} is the first spectrogonio-radiometer at IPAG. Especially designed for icy and bright surfaces, it consists of two rotating arms to change the incidence and the emergence angles. SHADOWS is based on SHINE, and comparison between the two instruments will be developed in this article.\\
\hspace*{0.5cm} The goniometer of the Bloomsburg University Goniometer Laboratory (BUG Lab) \cite{BUG,BUG-description} is similar to SHINE and SHADOWS, with its two rotating arms, allowing measurements from 0° to 65° in incidence, 0° to 80° in measurement and from 0° to 180° in azimuth. The monochromatic light source is a quartz halogen lamp passing through inteference filters, allowing measurements between 400 and 900 nm.\\
\hspace*{0.5cm} The Physikalisches Institute Radiometric Experiment PHIRE \cite{PHIRE} in Berne can conduct bidirectional reflectance measurements with a wide angular range. The light source is an quartz tungsten halogene lamp and the created beam passes through color filters to select the wavelength from 450 nm to 1064 nm. Its successor PHIRE-2 \cite{PHIRE2} has been especially designed to increase the signal-to-noise ratio and to work a sub-zero temperature. The light source is separated from the goniometer so it can be installed at room-temperature and the goniometer alone is placed in a freezer.\\
\hspace*{0.5cm} The Finnish Geodetic Institute’s field goniospectrometer FIGIFIGO \cite{FIGIFIGO} conducts in-situ reflectance measurements using the sunlight as light source, from 350 to 2500 nm. Optics look down to the target through a mirror on the top of the measurement arm. Optical fibers, spectrometer, control computer and all the electronics are contained in the casing on the ground.\\
\hspace*{0.5cm}The goniometer at the ONERA/DOTA at Toulouse, France \citep{onera1,onera2} is designed to conduct bidirectional reflectance measurements in the laboratory and in-situ. The light source is a quartz-halogen-tungsten lamp moveable to correspond to an incidence angle between 0° and 60°. The measurement camera and spectrometer can rotate from 0° to 60° in emergence angle, and from 0° to 180° in azimuth. The instrument covers the spectral range from 420 nm to 950 nm.

\section{Presentation of the instrument}
\begin{figure}[h!]
\includegraphics[scale=0.42]{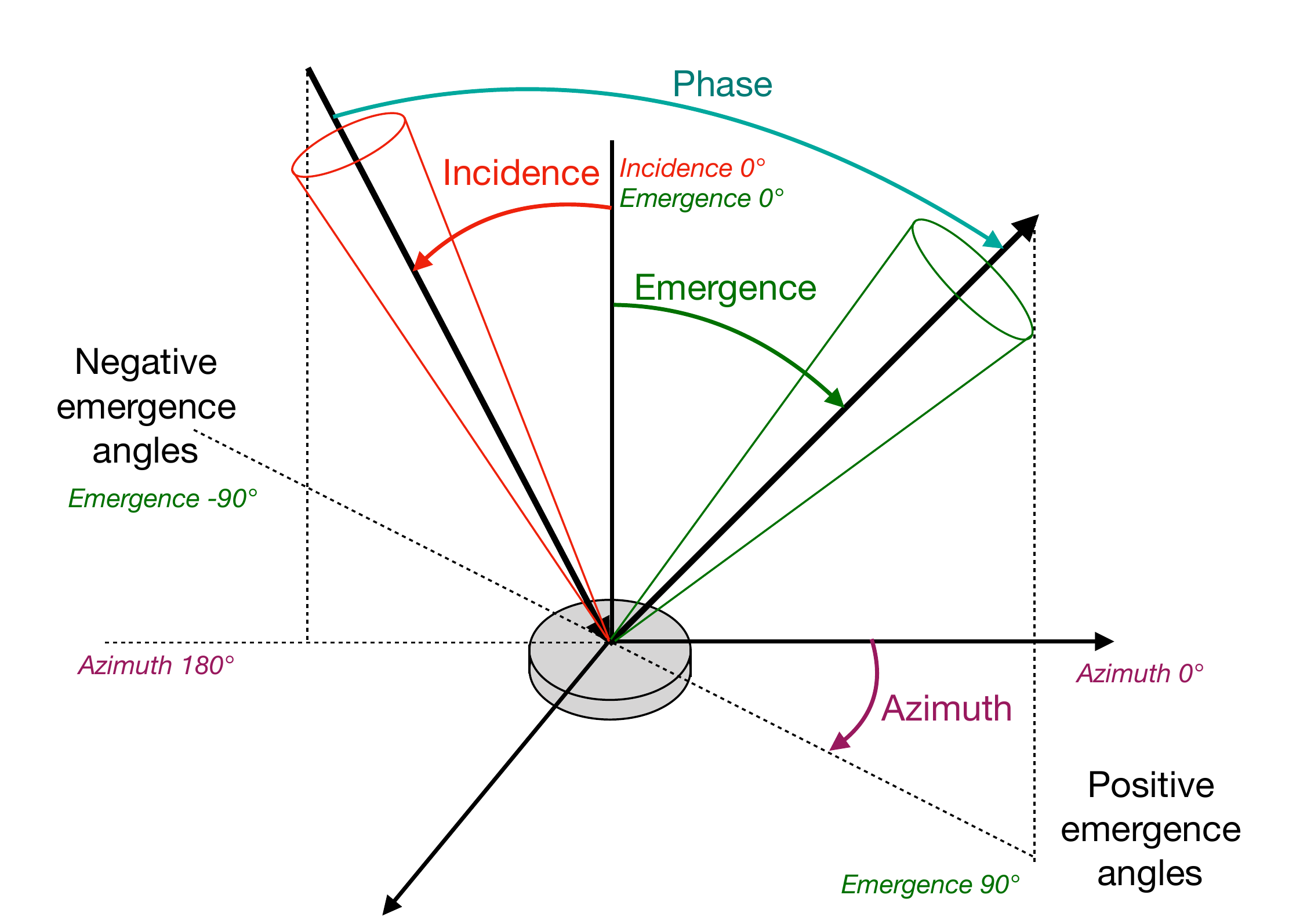}
\caption{Definition of the incidence, emergence and azimuth angles used for the reflectance measurements.}
\label{schema angles}
\end{figure}

\begin{figure*}[h!]
\begin{center}
\includegraphics[scale=0.7]{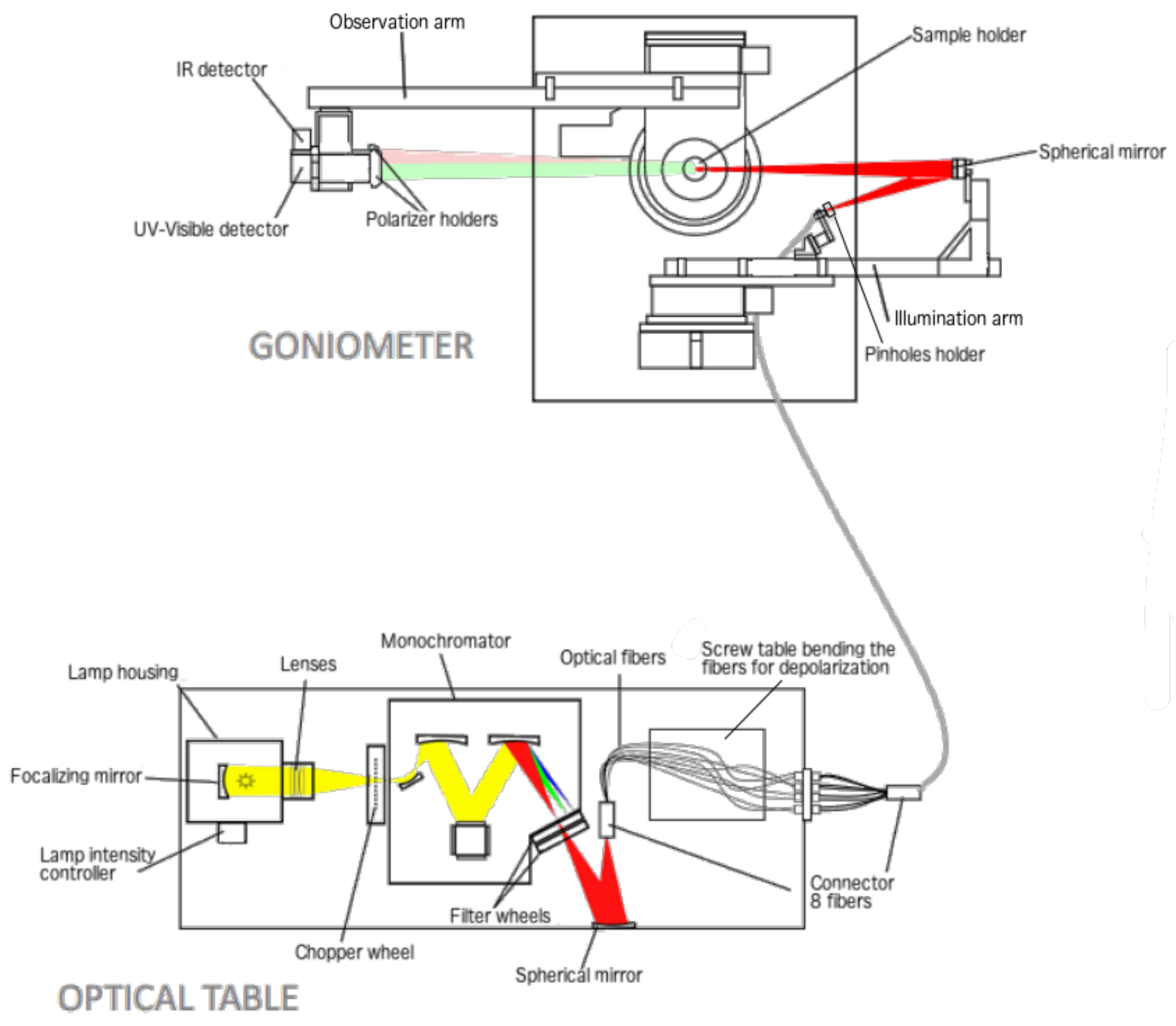}
\caption{Schematic view of SHADOWS, showing its two principal parts: the optical table where the monochromatic light is generated and the goniometer illuminating the sample and collecting the scattered light. The optical fiber bundle flexibly connects the two parts.}
\label{shadows schema complet}
\end{center}
\end{figure*}

The general design of SHADOWS is as follows: A monochromatic light is generated and scanned on an optical table and sent through a bundle of optical fibers to a mirror on the goniometer's illumination arm. The sample scatters the focused illumination beam and two detectors, located on the goniometer's observation arm, collect the light scattered in the UV-Visible and infrared ranges. The different angles, incidence, emergence and azimuth are explicited on figure \ref{schema angles} as they are defined for the goniometer. The choice to set the positive values of emergence angle around azimuth 0° and the negative values around azimuth 180° is arbitrary. Figure \ref{shadows schema complet} presents a complete scheme of the optical path of the instrument.

While the optical table is placed in the laboratory at room temperature, the goniometer is installed in a cold room with the optical fibers connecting the two parts of the instrument passing through the wall of the cold room. It is thus possible to study ices and other samples at a temperature as low as -20°C. The future cryogenic cell, and the vacuum chamber, will be placed in the center of the goniometer in place of the open sample holder.

\subsection{Optical table}
The light source is a 250W quartz-tungsten halogen commercial lamp (Oriel QTH 10-250W + OPS-Q250 power supply) placed in a housing equipped with a temperature-stabilized Silicon photodiode controlling and stabilizing the light intensity output to better than 0.1$\%$ peak-to-peak over 24h. The light is focused on the input slit of the monochromator with a homemade condenser (triplet of $CaF_2$ lenses) that also transmits infrared radiation. Just before the slit, the light is modulated by a chopper wheel at a frequency of 413 Hz, far from any perturbation coming from the 50Hz electrical network and its harmonics, and 100Hz cold room lamps and its harmonics. A 4-gratings monochromator (Oriel MS257) diffracts the incoming light and focuses it on the output slit using torus mirrors to remove any chromatic aberration. The monochromator and the other instruments characteristics are presented in table \ref{tableau configurations nominales}. Both monochromator input and output slits are motorized and controlled by the software. The instrument can thus adjust the width of the slits during a spectral scan to maintain a relatively constant spectral resolution over the entire spectral range. Behind the output slit of the monochromator, two wheels holding the high-pass filters remove high-order reflections and stray light. The light exiting the monochromator is focused by a sperical mirror on a custom made bundle of 8 $ZrF_4$ optical fibers (manufacturer: Le Verre Fluoré).\\

\subsection{Fibers}
The 8 optical fibers have two purposes: the first is to collect and transport in a flexible way the monochromatic light from the optical table to the illumination mirror of the goniometer, the second is to depolarize the incoming light. Two bundles in series have been designed to achieve these goals.\\
\hspace*{0.5cm}In the first bundle, the 0.76m-long fibers are vertically aligned at one end, then separated to be individually connected to the second bundle. The alignment of the 8 optical fibers matches the image of the monochromator slit by the spherical mirror (magnification of 0.5). Each of the 8 fibers are then individually bent with a moderatly strong curvature to induce more reflections at the core-clad interface, resulting in a strong depolarization of light. Achromatic depolarizers, such as the quartz wedge depolarizer \cite{quartz-depolarizer}, are not suitable for the instrument because their spectral range does not cover the full range of SHADOWS. The polarization of the incident light, and several other depolarizing options considered are described in section 5.C.4.\\
\hspace*{0.5cm}The fibers are individually connected to the 2m-long bundle, the output being arranged in a circle of 2 mm in diameter. This stainless steel sheathed bundle remains flexible at temperatures as low as -20°C. \\ 
\hspace*{0.5cm}The fibers have a core diameter of 600$\mu$m, which sets the maximum width of the output slit at 1.2mm and, therefore, the largest spectral bandwidth for each monochromator grating (see Table \ref{tableau configurations nominales}).\label{fibers}

\subsection{Illumination}
\begin{figure}[h!]
\begin{center}
\includegraphics[scale=0.5]{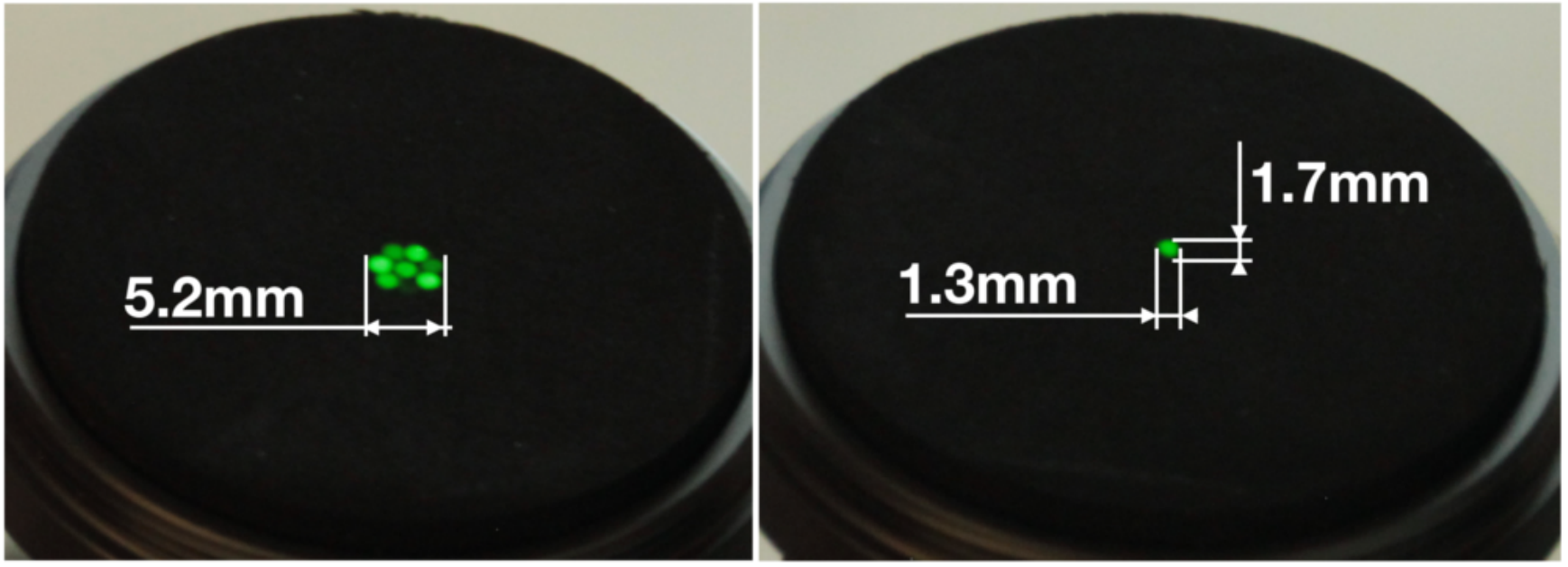}
\caption{Picture of the nominal illumination spot (left) and of the reduced illumination spot (right) of SHADOWS on a 2$\%$ Spectralon, seen from a emergence angle around 40°.}
\label{illumination shadows}
\end{center}
\end{figure}

The output of the fibers bundle is placed at the focal point of a spherical mirror, held by the illumination arm of the goniometer. With a diameter of 50.8 mm and a focal length of 220 mm, this mirror creates the image of the fibers on the sample (figure \ref{illumination shadows}), resulting in a nadir illumination spot of 5.2 mm in diameter, with a convergence half-angle of 2.9°. This value defines the angular resolution of the illumination. \\
\hspace*{0.5cm}The size of the illumination spot can be reduced by using a set of pinholes placed in front of the fibers output with a two-axis translation stage for better adjustements. In this configuration, light coming from only one fiber can exit, resulting in an illumination spot of 1.7mm by 1.3mm (figure \ref{illumination shadows}).

One of the fibers was partially broken during the installation of the goniometer and thus transmits less than 5$\%$ of the incoming light.\\
\hspace*{0.5cm} Due to a small part of the illumination being blocked by the translation stage, the reduced illumination spot tends to present an oval shape, rather than a perfect disc. For now, three pinholes are available, 500 $\mu$m, 600$\mu$m and 700$\mu$m, but the collection is to be extended in the future. Those three pinholes can let pass the light from at least one fiber, it is possible using the translation stage to partially mask a fiber and select half of its output or let pass the ligth from more than one fiber. This reduces or increases the illumination spot (table \ref{table pinholes}) as well as the signal-to-noise ratio.\\

\begin{table}[h!]
\begin{center}
\caption{Typical size of the reduced illumination spot letting the light from only one fiber pass, and maximum size of the spot letting the light from several fibers pass, for the three pinholes available for now.}
\begin{tabular}{ccc}
\hline
Pinhole & Illumination spot (1 fiber) & Maximum size\\
\hline
500$\mu$m & 1.73mm by 1.32mm & 2.33mm by 1.31mm\\
600$\mu$m & 1.77mm by 1.37mm & 2.50mm by 1.64mm\\
700$\mu$m &1.80mm by 1.43mm & 2.80mm by 1.94mm\\
\hline
\end{tabular}
\label{table pinholes}
\end{center}
\end{table}

\subsection{Observation}
The radiance of the sample is measured by two mono-detectors, held by the observation arm of the goniometer. The visible wavelengths are covered by a silicon photodiode with a spectral response from 185 to 1200 nm, while the infrared wavelengths are covered by an InSb photovoltaic detector, cooled at 80K by a small cryocooler (Ricor K508), with a spectral response from 800 to 5200 nm. We designed a set of achromatic lense triplets in front of each detector to reduce the field of view to 20 mm diameter at the sample surface, and at a solid angle FWHM of 4.1°. This solid angle defines the nominal angular resolution of the observation, but can be reduced with the use of diaphragms in front of the optics, resulting in an angular resolution of 3.3°, 2.5° or 1.6° according to the diaphragms in place. To optimize their transmission, the lenses are made of sapphire, $CaF_2$ and Suprasil, and are treated with $MgF_2$ coating to reduce reflections. The transmission of each set of lenses is greater than 90$\%$ over the entire spectral range.\\
\hspace*{0.5cm}The observation area (20 mm at nadir) is generally larger than the sample (with diameter of 7mm for spectra under the nominal geometry of incidence 0° and emergence 30°, or 7mm by 14mm for BRDF measurements) and much larger than the illumination spot (5.2 mm, or less, at nadir), whatever the geometry of observation (incidence, emergence, azimuth angles), so as to guarantee the collection of all the photons reflected in the direction of the detectors. The lock-in amplifiers of the synchronous detection remove all unmodulated background light scattered by the sample and the sample-holders, as well as direct and scattered thermal emission. They also automatically adjust their sensitivity based on the measured signal to optimize the signal-to-noise ratio over a wide range of signal and reflectance levels.\\

\subsection{Goniometer}
The rotation of the two arms of the goniometer are ensured by three rotation stages with stepper motors. The first two allow rotation in a common vertical plane of the illumination arm from 0° to 90°, and the observation from 0° to 90° on each side to the normal to the surface, while the latter one provides a horizontal rotation of 0° to 180° of the observation arm to change the relative azimuth between illumination and observation. The angular resolution of the stepper motors is 0.001°. \\
\hspace*{0.5cm}The torque of the motors is 60 Nm for the emergence and azimuth displacement, and 35 Nm for the incidence arm, which is sufficient to support respectively the mass of the detectors and the illumination mirror. 3D modeling of the goniometer on SolidWorks and CATIA indicates that the mechanical deformations due to the elasticity of the materials are maximal when the arms are close to the horizontal. The displacement induced for the illumination spot at the surface of the sample can reach 0.028 mm, and up to 0.121 mm for the center of the observation area. These offsets are negligible compared to the diameter of the illumination spot and the observation zone (<0.6$\%$). \\
\hspace*{0.5cm}The whole structure is anodized in black, however, according to the measurements we made with our SHINE spectro-gonio radiometer, this anodization does not significantly absorb the near-infrared wavelengths between 0.7$\mu$m and 2.75$\mu$m, and has strong specular reflections. So we decided to cover both arms with a black paint (Peinture Noire Mate RAL9005, Castorama) with a diffuse reflectance measured between 5$\%$ and 10.1$\%$ over the entire spectral range of the instrument. This paint strongly limits the possible parasitic reflections on the arms that can reach the detectors. \\

\begin{figure}[h!]
\begin{center}
\includegraphics[scale=0.35]{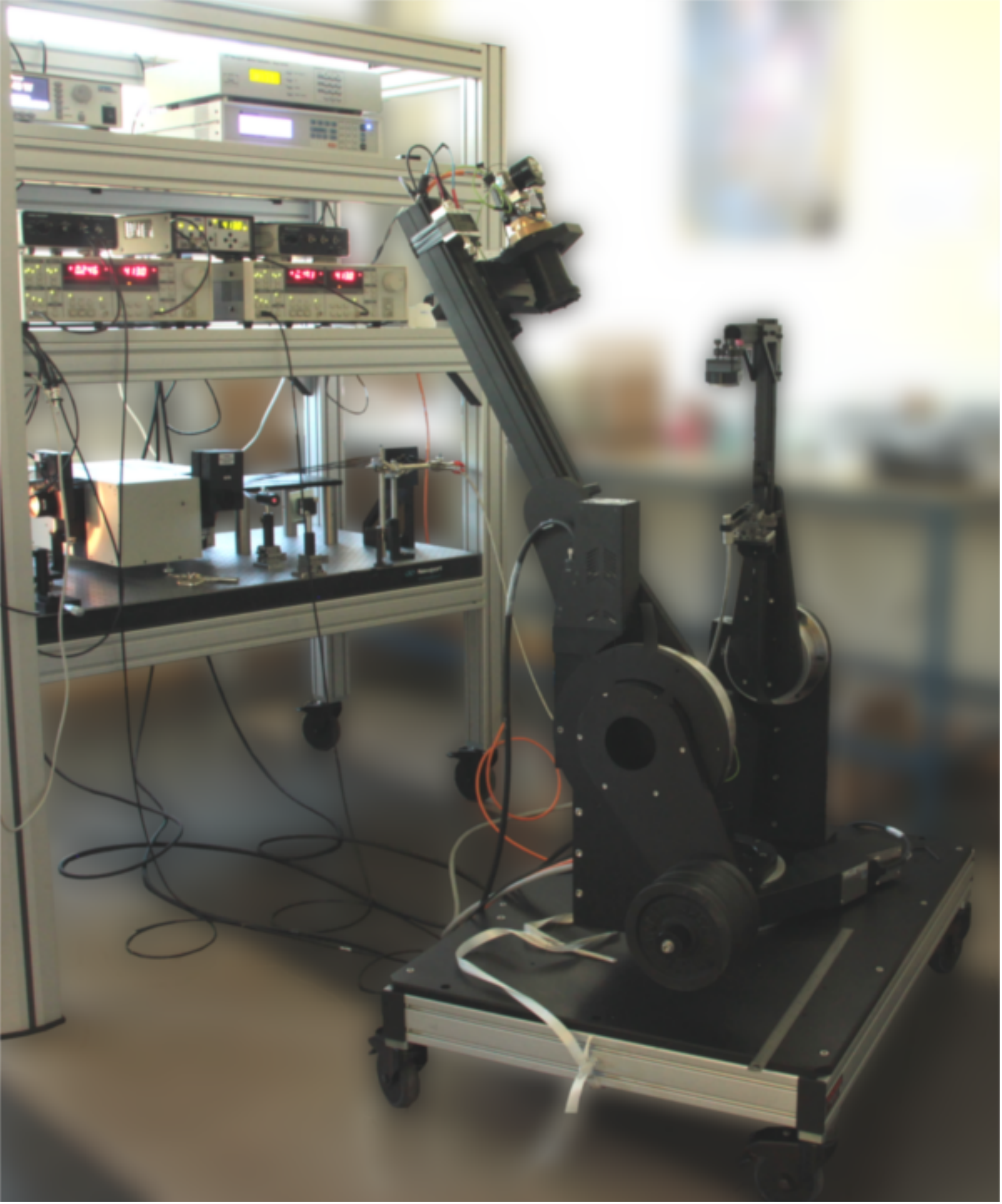}
\caption{The two main parts of SHADOWS: the optical table where the monochromatic incident light is generated and the goniometer with its 2 arms holding the optical fiber bundle and a mirror for the illumination of the sample (rear arm) and two detectors for the collection of reflected light (front arm). The instrumentation rack also contains the power and source stabilization, control of 3 rotation stages, detector amplifiers and lock-in amplifiers. The total height of the goniometer is 170 cm.}
\label{photo shadows}
\end{center}
\end{figure}

\subsection{Diffuse transmission}
\begin{figure}[h!]
\begin{center}
\includegraphics[scale=0.55]{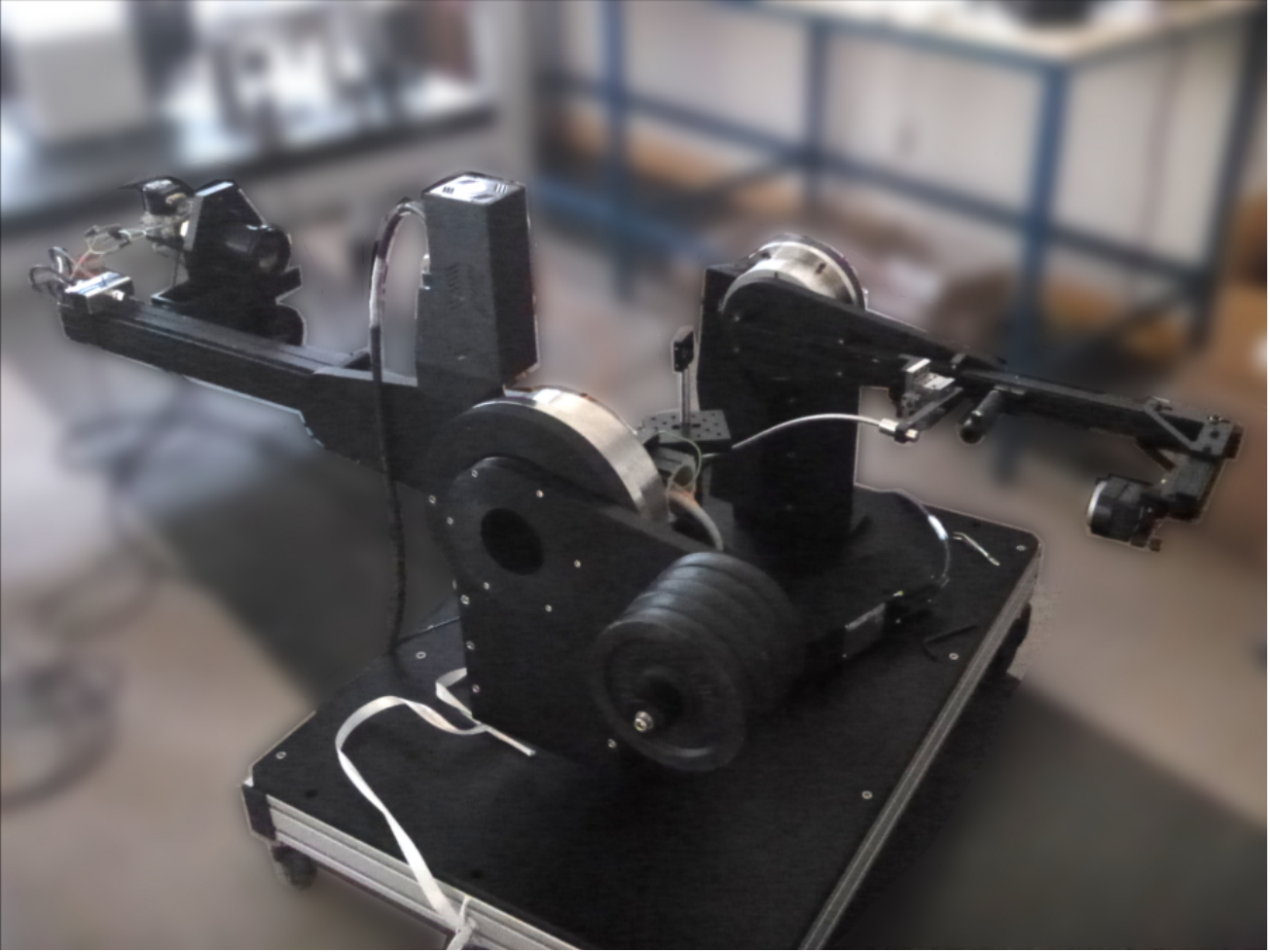}
\caption{SHADOWS goniometer in transmission mode. The sample is at the focal point of illumination.}
\label{photo shadows transmi}
\end{center}
\end{figure}

A diffuse transmission mode, that does not require modification of the setup, has been added. The arms are initially placed in a horizontal position, as shown in figure \ref{photo shadows transmi}. The sample is placed vertically at the focal point of illumination.\\

\hspace*{0.5cm}As for reflectance measurements, the illumination is focused on the sample with a half-angle of 2.9° and the detectors collect the transmitted light (after partial scattering in the sample) from an area of 20mm in diameter on the sample and in a half-angle of 2.05°. 
Simple diffuse transmission spectra can be measured with both incident and emergence arms fixed at 90° (incidence and emergence angles equal to 0°), but a complete characterization of the angular distribution of light scattered and transmitted by the sample can be performed by changing the emergence angle (up to 85° from the normal of the surface in the vertical plane). It is possible to explore in the horizontal planes at each emergence angles using the "azimuthal motor". Finally, it is also possible to vary the angle of incidence on the sample (up to 75°).\\
\hspace*{0.5cm}Measurements of transmission of non-diffusing materials is also possible. However the detector, 750 mm away from the sample, collects only part of the transmitted light in its acceptance angle. This may induce some photometric error in the transmission of thick crystals due to the refraction of the incident beam over a significant distance within the sample, which did not occur for the reference "white" measurements without sample. Thin samples will be prefered for transmission measurements with SHADOWS. \\
\hspace*{0.5cm}Since, in this case, the light is sent directly to the detectors, the measured intensity is much higher than in reflection mode. It is thus possible to reach spectral resolutions of less than 1 nm by reducing the slit widths of the monochromator. 

\subsection{Software}
A home-made control-acquisition software has been developed to control SHADOWS' instruments and define and calibrate the different types of measurements. This program fully controls the monochromator, goniometer, and lock-in amplifiers as well as some parameters of the future environment cells. The software can automatically calibrate, during acquisition, the raw measurements using reference spectra and then calculate the reflectance of the sample. It takes into account certain corrections related to the illumination-observation geometry, such as the modification of the size of the observation area with increasing emergence angle and the spatial response of both detectors (explicited in section 5.B), and performs the photometric calibration using previously measured BRDFs of the Spectralon and Infragold reference targets \cite{these-nicolas}. The photometric calibration used by the program is displayed by equation 1. The reference measurements are acquired at an incidence angle of 0°, an emergence angle of 30° and an azimuth angle of 0°.
\begin{equation}
{R_{sample}}^{(\lambda, \theta_{i}, \theta_{e}, \theta_{z})} = \frac{{S_{sample}}^{(\lambda, \theta_{i}, \theta_{e}, \theta_{z})} \cos(30^o)}{{S_{reference}}^{(\lambda,0^o, 30^o, 0^o)} \cos\theta_{e}} {R_{reference}}^{(\lambda, 0^o, 30^o, 0^o))}
\end{equation}

where ${R_{sample}}^{(\lambda, \theta_{i}, \theta_{e}, \theta_{z})}$ is the calculated bidirectional reflectance of the sample at wavelength $\lambda$ , incidence angle $\theta_{i}$, emergence angle $\theta_{e}$ and azimuth angle $\theta_{z}$, ${S_{sample}}^{(\lambda, \theta_{i}, \theta_{e}, \theta_{z})}$ the raw signal measured at $\lambda$, $\theta_{i}$, $\theta_{e}$ and $\theta_{z}$ on the sample, ${S_{reference}}^{(\lambda,0°, 30°, 0°)}$ the raw signal measured at wavelength $\lambda$, incidence angle $\theta_{ir}=0$°, emergence angle $\theta_{er}=30$° and azimuth angle $\theta_{zr}=0$° on the reference target, and ${R_{reference}}^{(\lambda, 0^o, 30^o, 0^o))}$ is the calibrated reflectance of the reference target at $\lambda$, $\theta_{ir}=0$°, $\theta_{er}=30$° and $\theta_{zr}=0$°. The Spectralon and Infragold targets are used as references.\\

\subsubsection{Options for flexible definition of parameters}
The software makes it possible to flexibly define the measurement wavelengths either on a single continuous spectral range, or by using several discrete ranges or even a discrete list of wavelengths. For each range or wavelength, the spectral resolution and the lower and upper limits of the signal-to-noise ratio (SNR) can be set. The same flexibility is found for defining the configuration of geometries where, for each of the angles of incidence, emergence and azimuth, one can define an entire range of angles, or only a series of specific geometries.\\ 
\hspace*{0.5cm}The monochromator holds 4 gratings to cover the entire spectral range of SHADOWS and 8 high-pass filters to remove stray-light from the monochromator. Spectral ranges of utilization of the filters are based on their different cutoff wavelengths and can be changed using the software in case of substitution. The wavelengths at which the monochromator changes the reflective grating can be set by the operator but standard configurations are pre-registered in the software. These configurations select the wavelengths at which the monochromator changes the grating according to the flux, spectral resolution or polarization of the light reflected by the gratings. Choice is given between the highest flux, the best spectral resolution, or the lowest polarization rate.\\
\hspace*{0.5cm}When preparing a BRDF measurement, the software analyses the list of requested geometries and remove "blind geometries" where mechanical parts of the goniometer itself blocks either the illumination light of the reflected light. This happens at phase angles lower than 5° when the detectors are above the spherical illumination mirror, or in special configurations outside the principal plane when light can be blocked by the motors. To reduce the number of time the goniometer goes near a dangerous zone, the software tries to reduce the number of movements of the arms, and favours small movements. The software creates a list of angular configurations in a precise order according to the following rules:
\begin{itemize}
\item Negative to positive azimuths, increasing illumination angle and increasing emergence angle.
\item Back and forth order if possible: if for example the measurement arm goes from -70° to 70° for one illumination configuration, the arm will goes from 70° to -70° for the next illumination angle. This rule also applies for the incidence arm in case of several azimuth angles.
\end{itemize} 
\hspace*{0.5cm}The software also prevents the goniometer going into any configuration that could be dangerous, such as near the optical fibers, walls or floor of the cold room, and crossing the two arms. The "safe zone" of the goniometer covers more than half of the hemisphere above the sample. The whole BRDF is constructed by using the principal plane as an axis of symetry for the scattering behaviour of the sample.\\
\hspace*{0.5cm}For experiments requiring measurements with constant spectral resolution, an option allows SHADOWS to be set to a fixed value over the entire spectral range. The software drives the motorized slits of the monochromator to adjust their width to maintain the spectral resolution around the desired value. \\
\hspace*{0.5cm}Another option allows dynamic optimization of the signal-to-noise ratio between the minimum and maximum values set by adjusting the time constant of the lock-in amplifiers. Finally, when the time constant is set to a long value, greater than 1 second, another option optimizes the acquisition time of a complete spectrum by reducing the time constant to 100 ms at each change of wavelength, and swtiching it back to the value fixed by the operator when the monochromator is set and the signal stable at the desired wavelength for the measurement. This option drastically reduces the acquisition time, which is particularly useful for spectra over the entire spectral range, or for BRDFs.\\

\subsubsection{Control of the environmental chambers}
The software also offers several types of methods to define a series of acquisitions, including one allowing the control of the future cryogenic cell and vacuum chamber. For example, the program can record the spectra of a sample placed in an environmental cell at a defined set of temperatures. After setting the temperature, the software monitors the temperature gauges of the cell until the sample's thermalization is complete, then starts the acquisition of a spectrum. At the end of the spectrum, the software sets the temperature controller at the next temperature in the series, waits for thermalization, starts a new spectrum, and so on. These series of temperature-dependent spectra allow the spectro-gonio radiometer to operate alone for several hours, or even days, without intervention of an operator. During an acquisition, the software registers the temperatures given by the temperature diodes in the cell, one placed near the heating resistor and the other on the sample holder, at each wavelength to get a monitoring over the duration of the spectrum, so around 40 minutes in the nominal configuration.

\section{Nominal configuration}
The standard configuration set for SHADOWS enables measurements of reflectance as low as 1$\%$. Most spectra are acquired using this configuration but modifications can be made to adapt the goniometer to the surface or its environment.\\
\hspace*{0.5cm}The nominal configuration for SHADOWS for the two types of measurement, reflectance and transmission, is described in table \ref{tableau configurations nominales}.\\

\begin{table*}[h!]
\begin{center}
\caption{Nominal configuration of SHADOWS for reflectance and transmission spectroscopy measurements. The minimum sampling of the incidence, emergence and azimuth angles correspond to the smallest motion of the stepper motors. The angular resolutions correspond to the half solid angle of the illumination and observation cones.}
\begin{tabular}{cc}
\hline
Characteristics & Quantities\\
\hline 
\multicolumn{2}{c}{\textit{SPECTRAL RANGE}} \\
Nominal range & 400 - 4700 nm\\
Low SNR ranges (factor of 100 lower)& 300 - 400 nm and 4700 - 5000 nm\\
$CO_2$ absorption band & between 4200 - 4300 nm\\
for opposition effect & 400 - 1700 nm \\
\multicolumn{2}{c}{\textit{LAMP AND LAMP HOUSING}} \\
Intensity stabilizer & 0.1$\%$ over 24h \\
Chopper frequency & 413 Hz \\
\multicolumn{2}{c}{\textit{MONOCHROMATOR}} \\
Input \& output slits & Height: 15mm, width: from 4 $\mu$m to 2 mm\\
Gratings & 1) 350 - 680 nm, 1200 lines/mm - Max resolution: 6.4 nm\\
 & 2) 680 - 1400 nm, 600 lines/mm - Max resolution: 12.8 nm\\
 & 3) 1400 - 3600 nm, 300 lines/mm - Max resolution: 25.8 nm\\
 & 4) 3600 - 5000 nm, 150 lines/nm - Max resolution: 51.3 nm\\
Wavelength accuracy & Gratings 1 and 2 : 0.2 nm, Grating 3: 0.4 nm, Grating 4: 0.6 nm\\
\multicolumn{2}{c}{\textit{BIDIRECTIONAL REFLECTANCE}} \\
Incidence angle & 0° to 75° (60° for bright samples)\\
& Resolution (solid angle of illumination): $\pm$2.9°\\
& Minimum sampling: 0.001°\\
Emergence angle & 0° to $\pm$85°\\
& Resolution: $\pm$2.05° (options: 0.8°, 1.25°, 1.65° but lower SNR )\\
& Minimum sampling: 0.001°\\
Azimuth angle & 0° to 180°\\
& Resolution: $\pm$2.05° (options: 0.8°, 1.25°, 1.65° but lower SNR )\\
& Minimum sampling: 0.001°\\
Phase angle & 5° to 160°\\
& for bright samples $\approx$ 8° to 140°.\\
Illumination spot size on sample & 5.2 mm (nadir) (option: 1.7 by 1.2mm or less, but lower SNR)\\
Observation FOV on sample & diameter 20 mm (nadir) \\
\multicolumn{2}{c}{\textit{TRANSMISSION}} \\
Incidence angle & 0° to 75°\\
& Resolution: $\pm$2.9°\\
Emergence angle & 0° to 85° in both H \& V planes (direct transmission at 0°)\\
& Resolution: $\pm$2.05° (options: 0.8°, 1.25°, 1.65° but lower SNR )\\
& Minimum sampling 0.001°\\
\hline
\end{tabular}
\label{tableau configurations nominales}
\end{center}
\end{table*}

For spectral analysis only, at fixed geometry, the spectra are acquired at nadir illumination and with an observation angle of 30°, with the lock-in amplifiers set to a time constant of 300ms. The spectral resolution can be fixed to 5 nm, but this value can be changed. In the nominal configurations, for spectra over the whole spectral range with a spectral sampling of 20 nm, the acquisition takes around 40 minutes. Changing the spectral sampling or the time constant modifies the acquisition time.

\section{Samples limitations}
\subsection{Texture of the surface}
The spectrogonio-radiometer has been designed to measure the reflectance of subcentimetric samples with small grain sizes. Typically, more than 100 grains must be illuminated at the surface for the measurement to be statistically relevant, so to ensure a wide variety of incidence angles on their facets and thus average the first external reflection (individual specular contribution). With the full illumination spot of 5.2 mm in diameter, the maximum grain size is about 500 $\mu$m or less for well crystalized samples. When using the pinhole that limits the illumination to one fiber, the maximum grain size is about 150 $\mu$m.

\subsection{Sample size}
The minimum size of the sample needed to obtain the maximum photometric accuracy is constrained by the illumination spot at the maximum incidence angle measured, the grain size and the reflectance of the sample. This constraint can be simply expressed as follow: all the incident photons (or at least a very large fraction, depending on the desired photometric accuracy) are only scattered or absorbed by the sample materials. This means that no photon should interact with the sides or the bottom of the sample holder. The minimum size of the sample should thus take into account, at least empirically using the Hapke scattering model, the maximum scattering length of the photons in a sample with a given reflectance, grain size, and porosity linked to the grain size distribution.\\
\hspace*{0.5cm}For dark samples, typically with a reflectance less than 0.2 over the whole spectral range, the lateral internal scattering of the light is strongly limited. The size of the sample must then be from 0.5 to 5 mm larger than the illumination spot, depending on the size of the grains. The required sample size is at its minimum for the nadir illumination. Also, to avoid having any contribution of the sample holder to the reflected signal, the sample depth must be at least 10 grains diameters. As an example, for nadir illumination, a dark sample with a grain size inferior to 25 $\mu$m and a 50 $\%$ porosity can have a minimum diameter of 5.7 mm and be 0.25 mm thick. This is equivalent to a volume of 6.5 ${mm^3}$. When reducing the illumination spot to the central fiber, the minimum sample diameter is reduced to 2 mm, and its volume to 0.8 ${mm^3}$, so of the order of a milligram of material. For micron-sized granular samples, this amount can be again reduced to less than 0.1 ${mm^3}$, so 100 $\mu$g while keeping the photometric accuracy. For a BRDF study of the sample, with varying incidence angles, only the size of the sample along the principal axis is to be adapted to the maximum incidence angle measured. The minimum volume is so increased by a factor of $1/\cos{\theta_{i max}}$. For example, for a maximum illumination angle of 60°, the previous dark sample is to be contained in a rectangular sample holder of at least 5.7 mm by 11 mm wide, and 0.25 mm deep. The minimum needed volume is then 16 ${mm^3}$.

\subsection{Bright samples}
Due to multiple diffusions inside the sample, the path length of the light is longer for a bright sample, compared to a dark one. Bright samples require wider margins around the illumination spot and greater thickness, both typically 100 times the grain size for a reflectance higher that 0.7. For example, for a maximum illumination angle of 60°, the sample holder for a bright sample with 25$\mu$m grain size should be at least 10.2 mm by 15.5 mm wide and 2.5 mm deep. This is equivalent to a volume of 400 $mm^3$.

\section{Reflectance calibrations on reference targets}

\subsection{Signal-to-noise ratio and reproducibility}
\begin{figure}[h!]
\begin{center}
\includegraphics[scale=0.9]{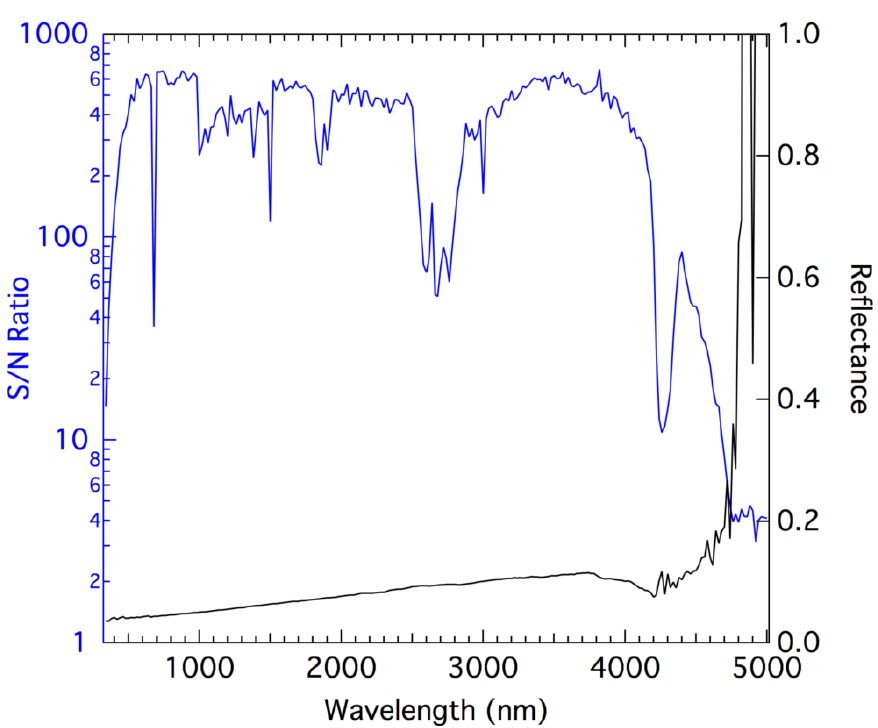}
\caption{Reflectance spectrum of the Spectralon 5$\%$ (black) and the associated signal-to-noise ratio (blue), represented by the measured reflectance averaged over 50 spectra of the Spectralon 5$\%$ and divided by the standard deviation. The spectra were acquired with a nadir illumination and an observation angle of 30°, and in the nominal configuration of 10 measurements at each wavelength with a time constant of 300 ms.}
\label{courbe SNR + reflec}
\end{center}
\end{figure}

The synchronous detection method using lock-in amplifiers removes the parasitic contributions of light and thermal backgrounds, and the InSb infrared detector is cryocooled at 80K to reduce its internal noise. But signal variations can come from the goniometer itself, as the use of 5 m-long BNC cables can alter the signal, and from the sensitivity and time constant of the signal detection. The control software adjusts the sensitivity at each wavelength to ensure the best measurement and signal-to-noise ratio, and the time constant of the lock-in amplifiers is set before the acquisition by the operator. A optional routine in the control software adjusts the time constant during the acquisition to keep the signal-to-noise ratio between two fixed values. \\
\hspace*{0.5cm}During an acquisition, at each wavelength, the control software records 10 measurements of the signal, separated by a time-lapse corresponding to the time constant of the lock-in amplifiers. The average value of these 10 measurements is taken as the measured signal at this wavelength, and its standard deviation is taken as the detection error. The number of measurements for the average, and the time constant can be changed by the operator. Figure \ref{courbe SNR + reflec} represents the signal-to-noise ratio taken over 50 reflectance spectra of the Spectralon 5$\%$.

\hspace*{0.5cm}Before 350 nm and after 4500 nm, the transmission of the optical fibers decreases radically, inducing a decrease of the signal-to-noise ratio. On spectra of very dark surfaces, the increase of reflectance at the end of the spectrum is an artefact due to a small bias on very low signals. The only way to reduce this bias is to increase the lamp intensity and so the amount of light travelling through the system.\\
\hspace*{0.5cm}The stabilization of the illumination lamp and the synchronous detection ensure a good reproducibility of the signal over time. But during long-time series of acquisitions, mostly over a couple of days, spectral variations can occur as reflectance peaks or absorption bands around 2700 nm and 4300 nm, respectively due to $H_2O$ vapor and $CO_2$ gas. Indeed the modulated light travels more than 300 cm through open air, first between the chopper and the entrance of the optical fiber bundle, where multiple reflections and the diffraction inside the monochromator increase the length of the light path, and then between the fiber output and the detector in the goniometer itself. So in these spectral ranges the composition of the air (water vapor and $CO_2$) in the cold chamber impacts the photometry caught by the detectors. Contents of water vapor and $CO_2$ in the atmosphere can vary in time according to the external humidity and also to the number of people around the goniometer. Their effects on the spectra can be compensated by regular measurements of reference targets. 

\subsection{Homogeneity of the observation}
\begin{figure}[h!]
\begin{subfigure}[c]{0.2\textwidth}
\includegraphics[scale=0.47]{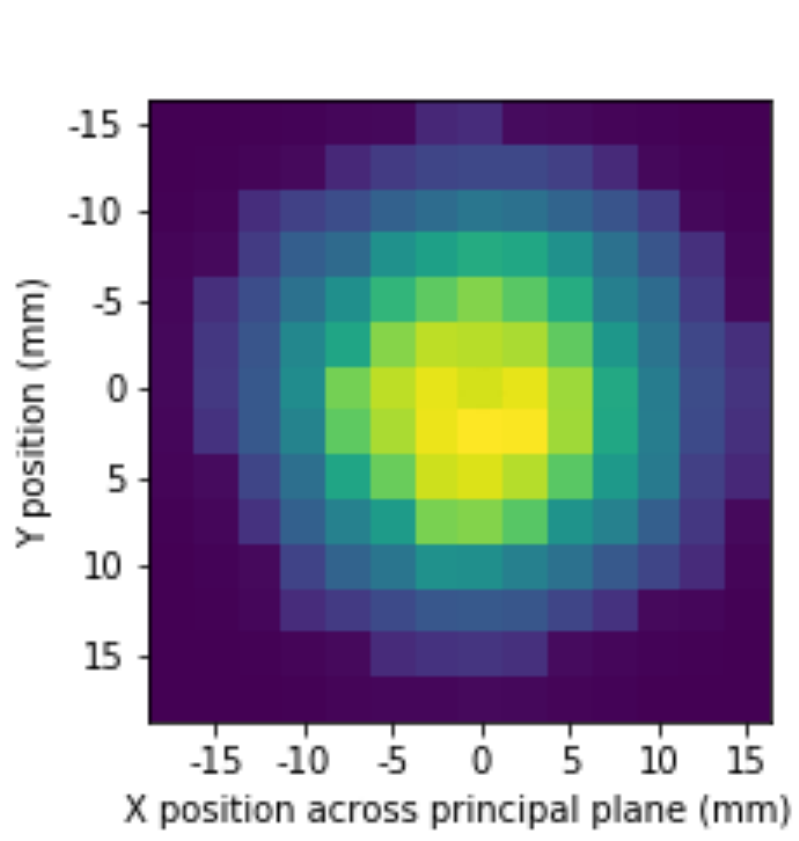}
\caption{Relative response of the visible detector at nadir observation (0°).}
\end{subfigure}~
\begin{subfigure}[c]{0.2\textwidth}
\includegraphics[scale=0.47]{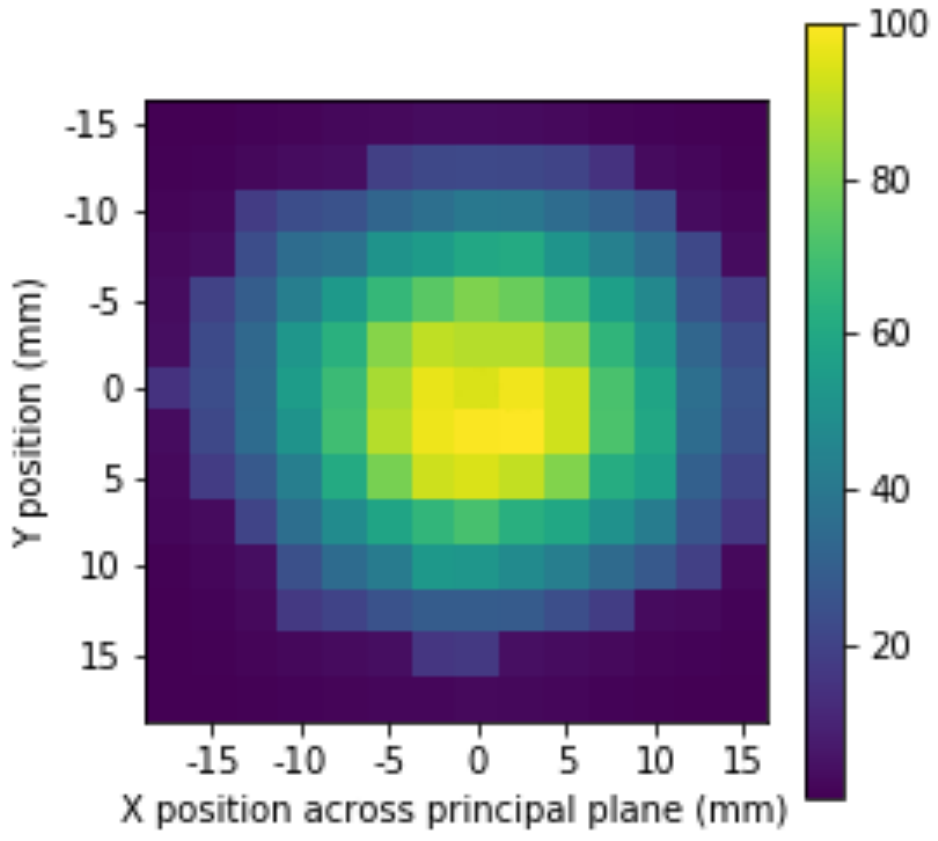}
\caption{Relative response of the visible detector at an observation angle of 30°.}
\end{subfigure}\\
\begin{subfigure}[c]{0.2\textwidth}
\includegraphics[scale=0.47]{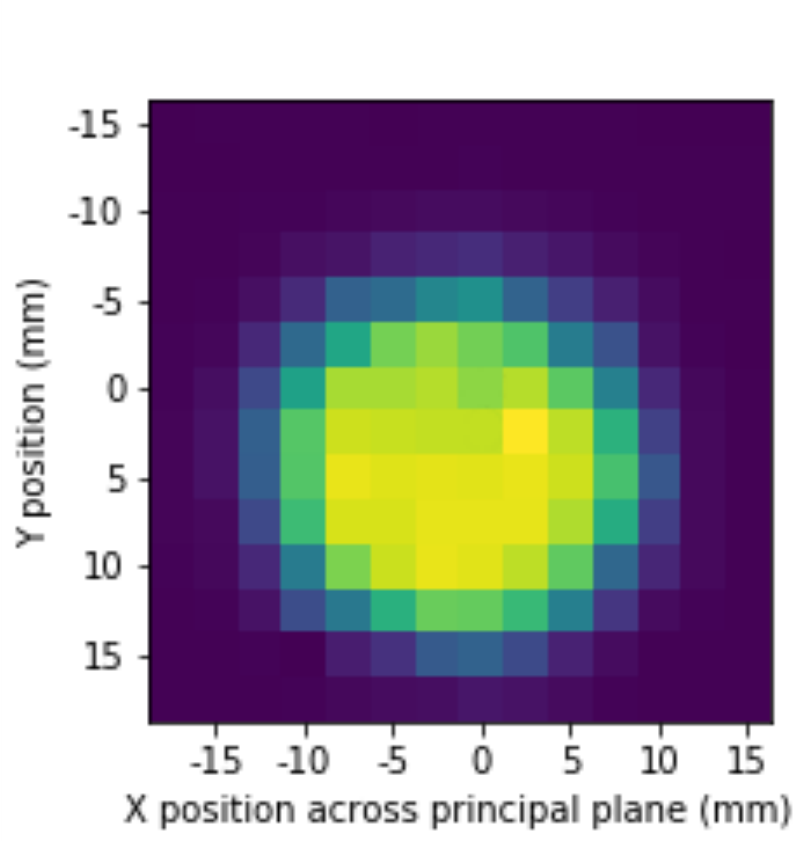}
\caption{Relative response of the infrared detector at nadir observation (0°).}
\end{subfigure}~
\begin{subfigure}[c]{0.2\textwidth}
\includegraphics[scale=0.47]{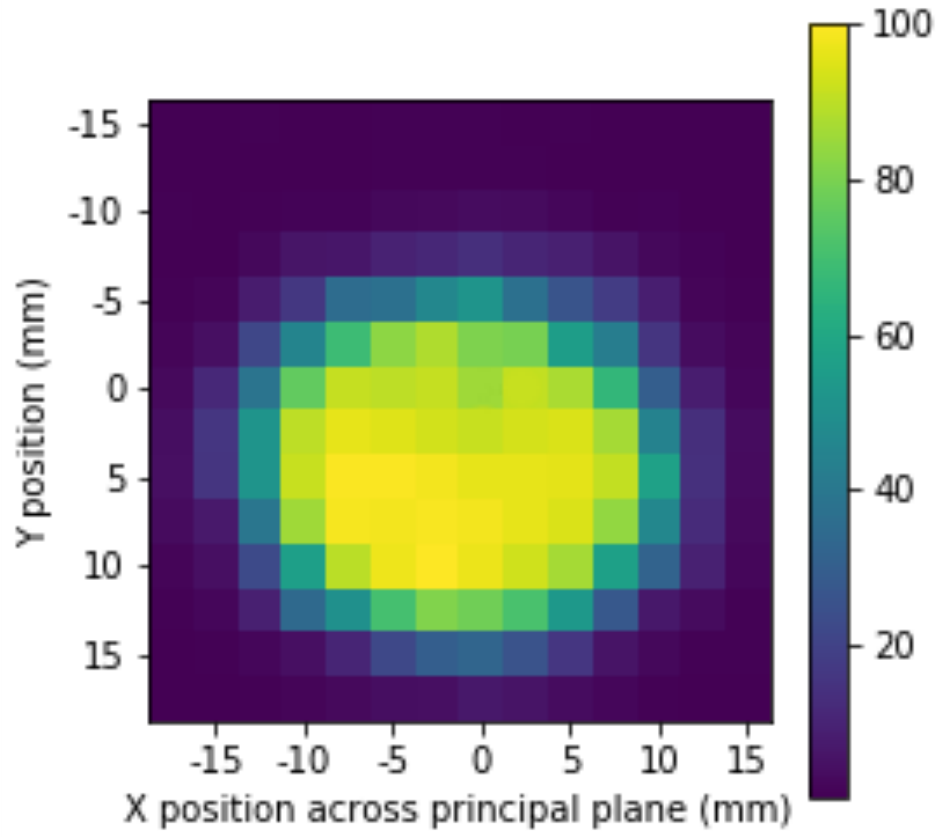}
\caption{Relative response of the infrared detector at an observation angle of 30°.}
\end{subfigure}
\caption{Homogeneity maps of the response of the two detectors, visible (top) and infrared (bottom) at nadir observation (left) and an observation angle of 30° (right). Each pixel corresponds to a measurement. The left and top scales are in mm. The color scale represents the relative intensity measured for each detector.}
\label{cartes homogen detectors}
\end{figure}

Due to its size, the illumination spot is always contained in the observation area of the two detectors, whatever the geometry. The size of the illumination spot depends on the angle of illumination with a cosine relation. The observation zone corresponds for each detector to a 20 mm-wide disk around the sample, at nadir observation. The observation zone becomes an ellipse at larger observation angle. To visualize the real observation zones of the detectors, a serie of acquisitions has been performed using the reduced illumination spot. The spot is moved on the surface using the position screws on the mirror holder and the reflected light is caught by the two detectors at two different observation angles, 0° and 30°. The wavelength is set at 900nm for the visible detector then at 1100nm for the infrared detector. Measurements are acquired every 2.5 mm on a 35 mm by 35 mm grid (figure \ref{cartes homogen detectors}).\\

\hspace*{0.5cm}Table \ref{table reponse detect} describes the spatial response of the detectors according to the emergence angle.\\

\begin{table}[h!]
\caption{Size of the response zones of the two detectors at several observation angles.}
\begin{center}
\begin{tabular}{lll}
\hline
Detector & Response>90$\%$ & Response>80$\%$\\
\hline
Visible (nadir)& Diameter 10 mm & Diameter 15 mm\\
Infrared (nadir)& Diameter 15 mm & Diameter 17 mm\\
Visible (30°)& Ellipse 12 by 10 mm&Ellipse 15 by 12 mm\\
Infrared (30°)& Ellipse 20 by 15 mm&Ellipse 22 by 17 mm\\
\hline
\end{tabular}
\label{table reponse detect}
\end{center}
\end{table}

\hspace*{0.5cm}At nadir illumination, the size of the illumination spot is 5.2mm in diameter. This spot gets an oval shape with the size of its semi-major axis increasing with the cosine of the incidence angle. At an incidence angle of 30°, the semi-major axis of the spot gets to 6mm, and to 10.4mm at an incidence angle of 60°. For an incidence angle of lower than 60° and for an observation in the principal plane, the illumination spot is always contained in the observation areas with over 90$\%$ relative intensity, for both detectors. At wider illumination angles, the size of the spot exceeds the size of the sensitive observation areas and corrections have to be taken into account in the measurement of the reflected intensity.\\
Outside the principal plane, the observation ellipses cross the illumination spot, until the semi-major axis of the observation ellipses get perpendicular to the semi-major axis of the illumination spot at an azimuth of 90°. For illumination angles lower than 60°, the illumination spot is still contained in the sensitive observation areas, even at an azimuth angle of 90°. At wider illumination angles, the illumination spot is larger than the semi-minor axis of the obsevtion areas. The program calculates the proportion of the illumination spot contained in the field of view of the detectors and extrapolates the measured signal to the full illumination.

\subsection{Sources of photometric errors}
\subsubsection{Loss of signal and stray-light}
\begin{figure}[h!]
\begin{center}
\includegraphics[scale=0.6]{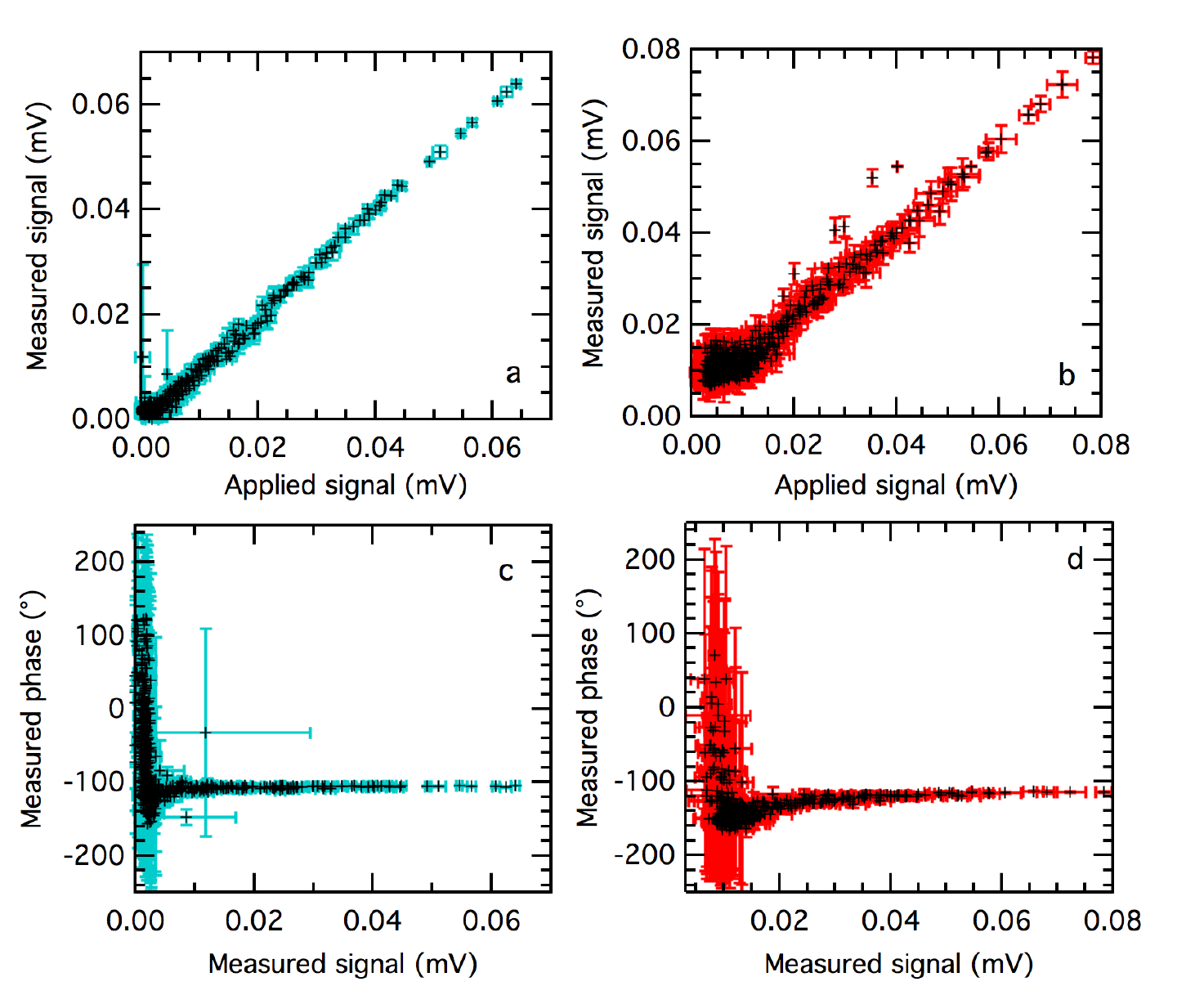}
\caption{Measured signal and phase for the two detectors at low signal. a: measured signal for the visible detector. b: measured signal for the infrared detector. c: measured phase for the visible detector. d: measured phase for the infrared detector.}
\label{non linéarité}
\end{center}
\end{figure}

With dark samples, the intensity of light reflected by the surface is only a few percents of the total incident flux. As the detectors only collect the monochromatic light reflected in a small solid angle ($\sim$0.06$\%$ of the hemisphere), they must be able to accurately measure very weak signals. Stray light coming from the monochromator can induce an offset, the detectors may be out of their linear response zone and the lock-in amplifiers may lose the modulation of very weak signals embedded in the thermal and light backgrounds. We ran tests to determine the non-linear response zone of the detectors as well as the limit of detections of the lock-in amplifiers. The results are displayed in figure \ref{non linéarité}.\\

\hspace*{0.5cm}At very low signal level the unstability of the phase measurement indicates a loss of the modulated signal by the lock-in amplifiers. No precise photometric measurements can be made if the lock-in amplifiers have lost the signal, and a signal offset usually results, as is clearly seen for the infrared channel in figure \ref{non linéarité}. To ensure accurate measurements, the detector amplifiers and the incident light level must be set so that the measured intensity is higher than the "loss of signal" threshold over most of the spectrum. 
On the other side, our measurements show that for all levels of signal, we are in the linear response zone of the detectors (fig. \ref{non linéarité}).

\subsubsection{Fibers curvature}
\begin{figure}[h!]
\begin{center}
\includegraphics[scale=0.45]{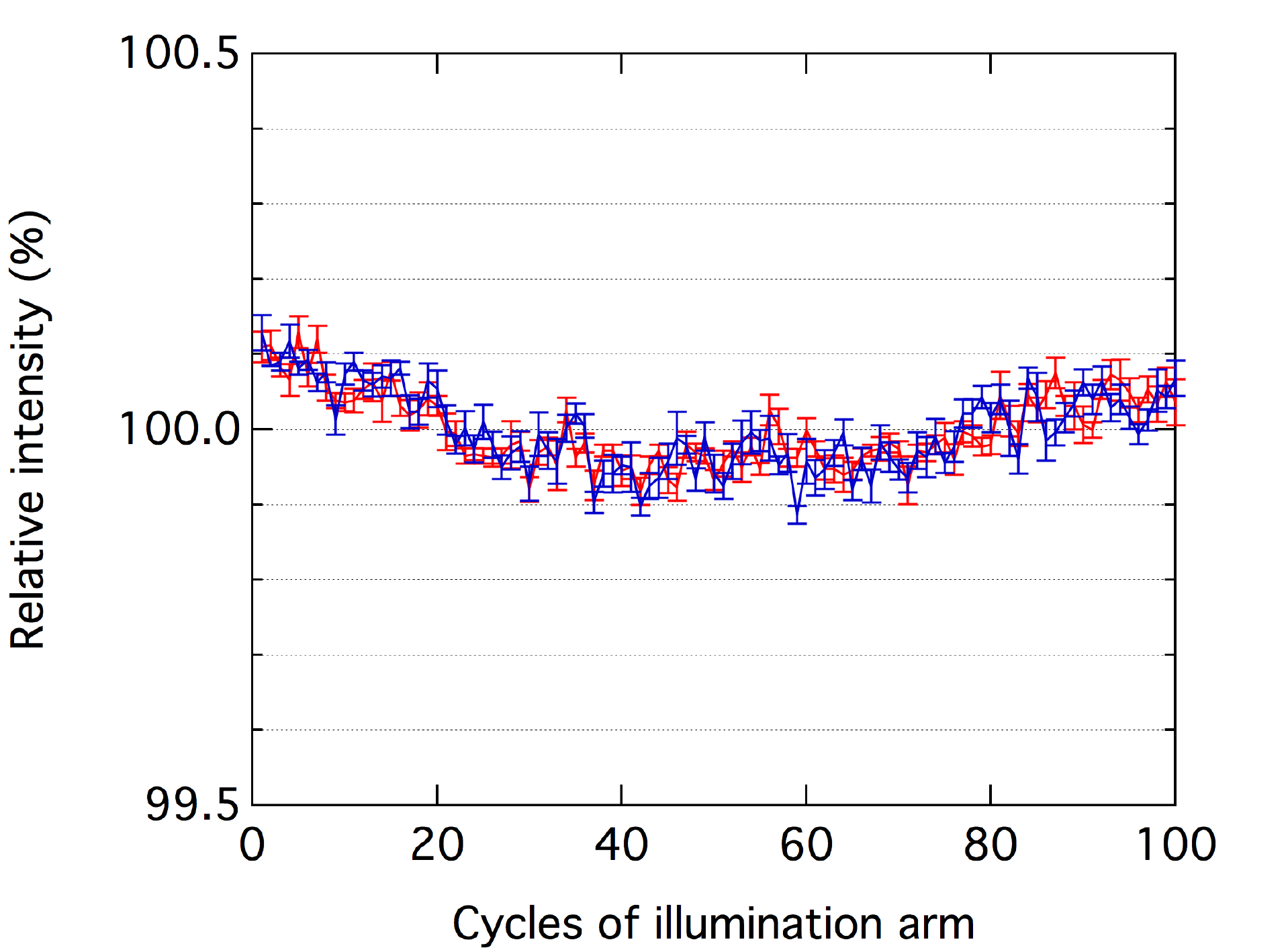}
\caption{Variations of intensity due to the change of the illumination angle between 0° (blue) and 60° (red), inducing repetitive variations in the curvature of the fiber bundle. The measures has been normalized by the average value of the whole serie. The general derive of the intensity corresponds to the quartz-halogen lamp outside of the chamber and so dependant on the thermal variations in the laboratory. Note that the whole range of the y-axis covers only 1$\%$.}
\label{hysteresis fibre}
\end{center}
\end{figure}

Changing the position of the illumination arm, i.e. the angle of incidence, induces a change in curvature of the fibers. These movements can induce variations in the flux at the exit of the fibers and therefore non-reproducible and hysteresis effects on the photometry. A series of 100 movements from 0° to 60° and back was imposed on the illumination arm and the reflectance of a spectralon target is measured at each incidence angle at a constant emergence of -30°. The cycles were separated by a waiting time of two-minutes, spreading the measurements over more than 5 hours. The results are presented in figure \ref{hysteresis fibre}. \\

A faint non-reproducibility with maximum relative variations of 0.21$\%$ at 0° and 0.24$\%$ at 60° is observed. The variation of signal due to the non reproducibility of the position of the arm has been measured to less than 0.1$\%$. The intensity of the lamp fluctuates over several hours but the variations have been proved to be lower than 0.1$\%$. The variation of intensity observed in figure \ref{hysteresis fibre} is due to the slow fluctuation of the temperature in the lab stabilized at 20°C. \\
\hspace*{0.5cm}Whatever the illumination angle is, the fibers bundle is never in contact with the goniometer or any other parts that could block the free movement of the bundle.

\subsubsection{Temperature variations}
\begin{figure}[h!]
\begin{center}
\includegraphics[scale=0.95]{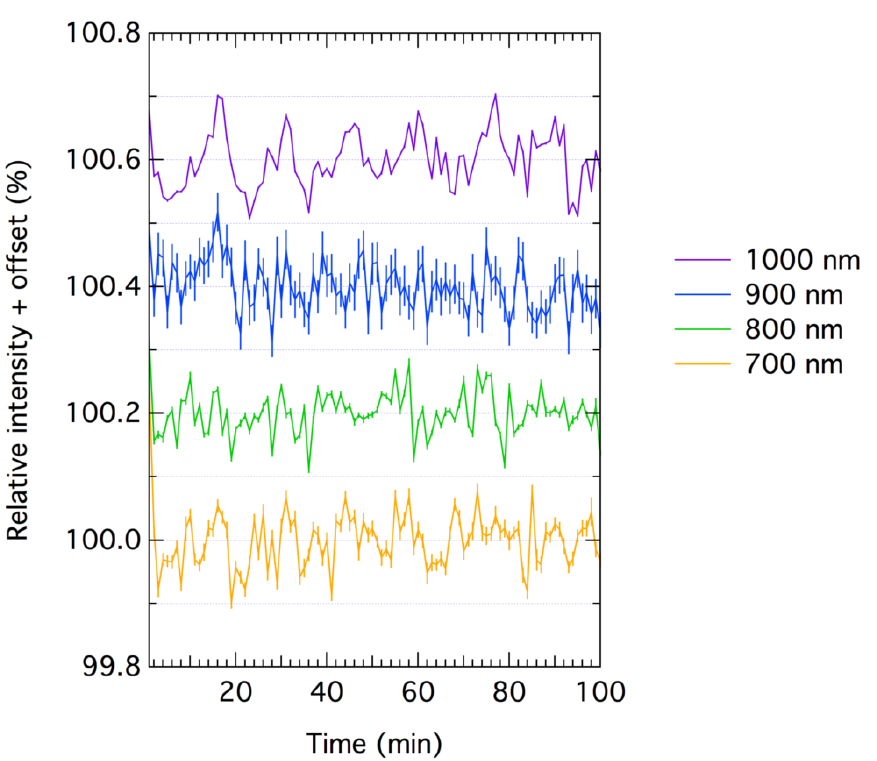}
\caption{Photometric variations of the Si visible detector acquired during around 100 minutes, so 6 thermalization cycles of the cold room (maximum temperature fluctuation : 3°C). Offset for clarity.}
\label{variations photo silicium}
\end{center}
\end{figure}

The spectral response of the visible Si photodiode is temperature dependant and can induce photometric errors. During a cooling cycle of the cold chamber, its temperature varies by 3°C over about 15 minutes. Figure \ref{variations photo silicium} shows the photometric variations of the silicium detector at different wavelengths. \\

\hspace*{0.5cm}The maximum variations occur at 1000 nm and represent 0.15$\%$ of the signal, so a variation of 0.05$\%$/°C. The constructor specifications of this detector declares increasing sensitivity to temperature with increasing wavelength with maximum variations of 0.6$\%$/°C at 1000nm. The mechanical parts and optics surrounding the detector add thermal inertia and so reduce the photometric errors due to temperature variations of the ambient air. The InSb infrared detector is insensible to these variations thanks to its cryocooler.\\

\subsubsection{Polarization of the illumination}
\begin{figure}[h!]
\begin{center}
\includegraphics[scale=0.87]{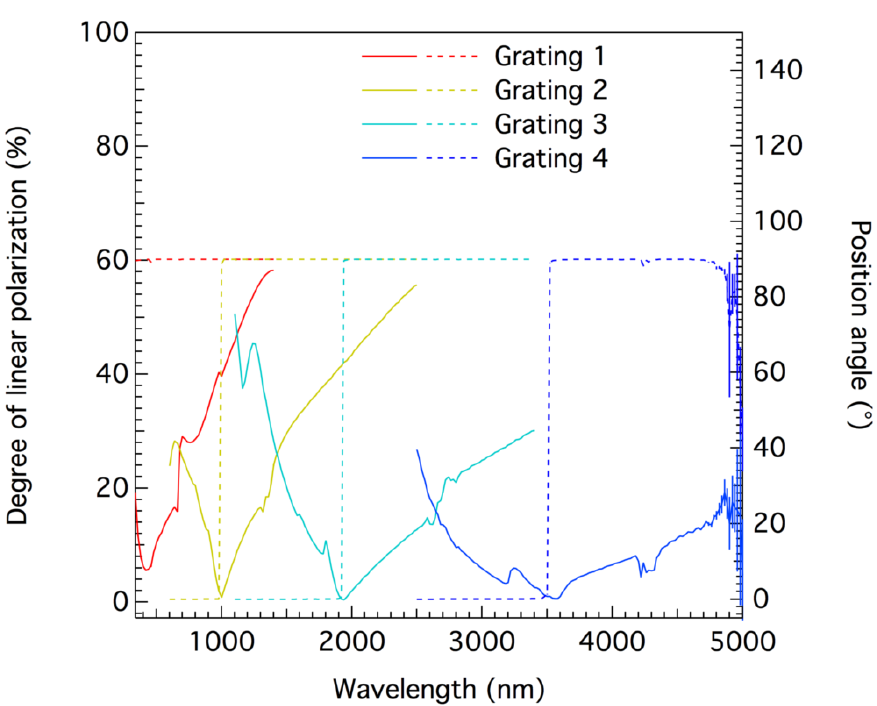}
\caption{Degree of linear polarization (solid lines) and position angle (dotted lines) of the incident light for the four gratings of the monochromator. The 0° of the polarizer was set parallel to the principal plane. The wavelengths at which we set the change of the monochromator gratings correspond to the intersections of the solid lines.}
\label{polar incidente}
\end{center}
\end{figure}

The first purpose of SHADOWS is to compare laboratory reflectance spectra with in-situ spectra of comets and asteroids, where the incident light is the sunlight, completely depolarized. Meteorites, terrestrial analogues or any artifical surfaces can present different reflectance behaviours according to the state of polarization of the incident light \cite{polar-spectralon}. The incident light of SHADOWS is partially polarized due to multiple reflections inside the monochromator and to the natural polarization rate of the halogen lamp around 7$\%$, but to remove any polarization effect on the measurements, the incident light has to be completely depolarized. The goal is to obtain a degree of linear polarization at the output of the fibers less than 0.1$\%$ over the whole spectral range to remove any small photometric variations due to the polarization of the incident light. This may be achieved by force-bending the optical fibers to force light refraction at the core-clad interface and so mix the different polarizations, as described in section \ref{fibers}. Figure \ref{polar incidente} represents the degree of linear polarization of the incident light for the four gratings of the monochromator without any constraints on the fibers. The measurements were made in transmission mode by vertically placing a grid polarizer on the sample holder and adjusted in height in order to make the illumination light pass throught the polarizer. Spectra were acquired for four different angles of the polarizers, 0° being parallel to the principal plane, 45°, 90° and 135°, in order to calculate the Stokes parameters Q, U and I and so deduce the degree of linear polarization as well of the position angle of the light \cite{stefano}. This calculation is registered in a polarimetry routine in the software and is not described in this article.

\begin{figure}[h!]
\begin{center}
\includegraphics[scale=0.9]{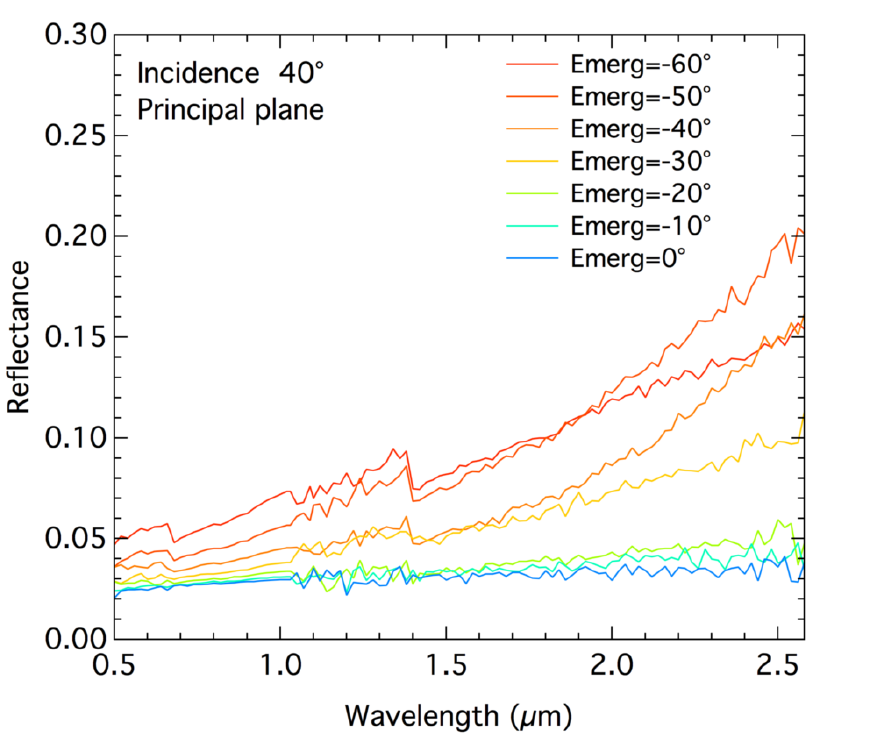}
\caption{Reflectance spectra of a black paint for several backscattering geometries. The effect of the incident polarization can been seen on 4 spectra as negative offsets at 0.68$\mu$m and 1.4 $\mu$m, with modification of the spectral slope.}
\label{effet polar}
\end{center}
\end{figure}

\hspace*{0.5cm}Dependance of the illumination polarization has been seen on the reflectance spectra as positive or negative offsets at wavelengths at which the monochromator changes the grating (with a change of position angle from 90° to 0°), as well as the modification of the slope of the continuum (see figure \ref{effet polar}). The grains of the sample can present different scattering behaviours for both polarizations, parallel and perpendicular to the scattering plane, usually called P and S polarizations. The polarization of the light scattered by each grating of the monochromator is polarized perpendicular to the principal plane of the goniometer, the 0° of the polarizer, so S polarized. With increasing wavelength, the degree of linear polarization decreases until it reaches a minimum where the angle of polarization turns to 90° and becomes perpendicular to the principal plane, so P polarized. The degree of linear polarization then increases until the ultimate wavelength possible for the grating. 

\hspace*{0.5cm}The modification of the slope of the continuum on the spectra is due to the variations of the degree of linear polarization. In addition, at each wavelength at which the monochromator changes the grating, the position angle passes from 90° to 0°, and creates offsets at these wavelengths on the reflectance spectra. \\
\hspace*{0.5cm}A first attempt of depolarization with the optical fibers under constraints (small curvature radius) turned out to be inefficient on the light from the monochromator focused on the fibers input, but removed almost 93$\%$ of the polarization of a fully polarized collimated 531.9 nm laser. After defocusing the injection of the light from the monochromator into the fibers, the degree of linear polarization at the position of the sample was measured with a maximum value of 2$\%$ and an average value of 1.3$\%$ over the whole spectral range. But this solution implies to lose roughly $90\%$ of the light at the injection.\\
\hspace*{0.5cm}The difference between the significant depolarization of the laser and the inefficiency with the output of the monochromator is thought to come from the amount of excited modes into the fibers. The laser being collimated with one degree of linear polarization, a small amount of modes are excited in the fibers when the light is injected. Constraints on the fibers will mix the light and depolarize it. But the partially polarized light getting out of the monochromator is focused on the fibers, so injected in a solid angle with a Numerical Aperture of 0.23. A lot of modes are already excited at the entrance of the fibers, and thus constraints on the fibers cannot mix the light more, resulting only in light losses.\\
\hspace*{0.5cm}To reach the goal of 0.1$\%$ over the whole spectral range, modifications have to be done to reduce the solid angle of injection, in addition to the curvature constraints applied to the fibers. The variation of the photometry of surfaces due to the incident polarization seen on reflectance spectra shows that the depolarizing setup has to be set permanently on the instrument. The chosen solution must present the highest depolarization rate with the lowest light loss. Several options are considered, each with their pros and cons. Diffusers are generally good depolarizers, especially fast rotating diffusers, but their transmission rates usually around 30$\%$ are too low for our setup. The same can be said about polarizers. Integrating the measurement over a full rotation of a fast-rotating polarizer remove the contribution of the input polarization, but making the light pass through a polarizer removes more than half the flux. It is also possible to reduce the angle of injection of the light on the input of the fibers using a set of achromatic lenses or mirrors. This solution can present the lowest absorption rate, but can modify the size of the image of the output of the monochromator on the input of the fibers, and so lose some light at the injection. This solution can be selected if the impact on the size of the image at the imput of the fibers does not implies a significant loss of light. The last solution would be to replace the multi-mode fibers under constraints by a bundle of octagonal fibers \cite{octagonal_fibers}. The capacity to scramble the light is higher than simple multi-mode fibers and are often used for high-precision spectrometers such as SPIRou \cite{spirou} at the CFHT, SOPHIE \cite{sophie,sophie+} at the 1.93-m OHP telescope or CHIRON \cite{CHIRON} at the 1.5m telescope at CTIO. The capacity of octagonal fibers to depolarize the light is to be tested with the optical table of SHADOWS.\\ The depolarization of the incident light is still a work in progress and compromises between the depolarization and the resulting light flux will be necessary. \\

\subsection{Cross-calibration}
\subsubsection{Reflectance spectra of calibration targets}
\begin{figure}[h!]
\begin{center}
\includegraphics[scale=0.9]{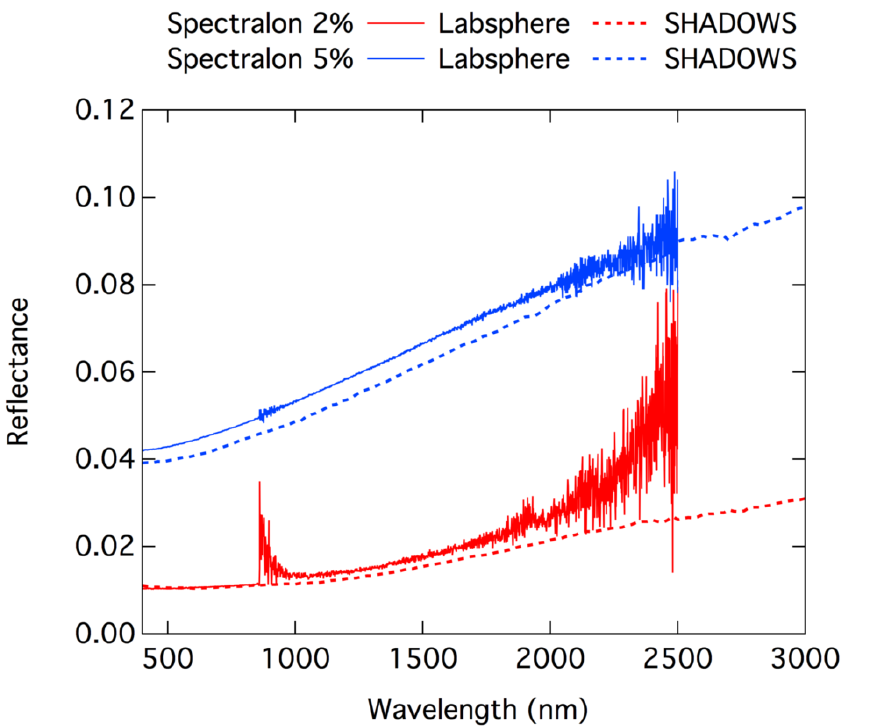}
\caption{Comparison between the calibrated directional-hemispheric spectra (illumination angle: 8°) of the Spectralon 5$\%$ reflectance target (blue) and the Spectralon 2$\%$ reflectance target (red) (data provided by Labsphere: solid lines), and the corresponding spectra measured by SHADOWS (dotted lines) with nadir illumination and an observation angle of 30°.}
\label{Compar spectralon 5 labsphere}
\end{center}
\end{figure}

The calibrated reflectance of reference targets, such as Labsphere's Spectralons, can be used to calibrate the spectral photometry of the instrument, as long as the types of reflectance can be compared. However, Labsphere uses directional-hemispheric reflectance with an incidence angle of 8° (in an integrating sphere), while SHADOWS is a bidirectional reflectance goniometer, so differences in photometry may appear as displayed in figure \ref{Compar spectralon 5 labsphere}. 
 
\hspace*{0.5cm}Reflectance spectra acquired with SHADOWS at i=0°, e=30° tend to show a lower reflectance value compared to the calibrated data. This difference has several origins. First SHADOWS measure bidirectional reflectance spectra and the non-lambertian behaviour of these targets should be taken into account.\\
\hspace*{0.5cm}Second, the calibration spectra provided by Labsphere are directional-hemispherical measurements (i=8°) in an integrating sphere. Bonnefoy \cite{these-nicolas}, using a full integration of BRDF measurements (on SHINE) of a Spectralon 99$\%$ reference target under the same illumination, showed that measurements in an integrating sphere overestimate by a few percents the true directional-hemispherical reflectance due to a lower contribution of the large angles in favor of the low phase angles with higher reflectance.\\

\subsubsection{Cross-calibration with SHINE spectro-gonio radiometer }
\begin{figure}[h!]
\begin{center}
\includegraphics[scale=0.9]{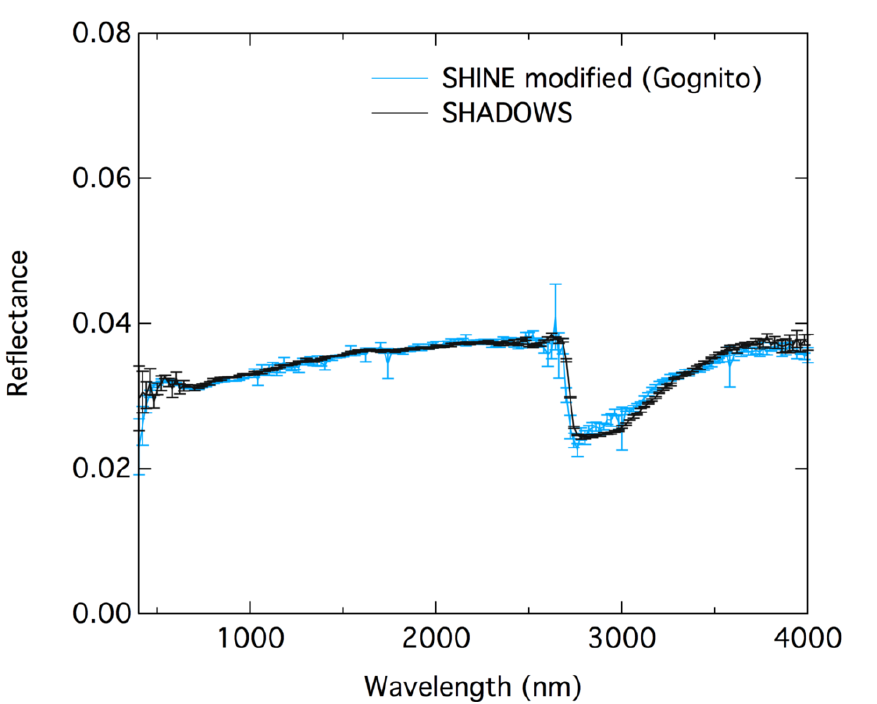}\\
\caption{Comparison of reflectance spectra of the dark Mukundpurar meteorite measured with SHADOWS (black) and the modified version of SHINE (blue) at incidence 0° and emergence 30°. The error bars are drawn on the spectra.}
\label{jaipur}
\end{center}
\end{figure}

SHINE has been fully calibrated by N. Bonnefoy during his PhD, and the photometric calibration processes and results are described in \cite{these-nicolas} and \cite{article_shine}. SHINE is used as a reference to check the photometric accuracy and calibration of SHADOWS. For dark surfaces, SHINE can be modified to simulate the illumination of SHADOWS by installing a spherical mirror at the ouput of the fibers. This optical variant of SHINE, nicknamed Gognito, has been used as a proof of concept for SHADOWS and enables measurements of reflectance lower than 0.15. It results in a wider illumination spot, 6mm in diameter, and a lower total light intensity than SHADOWS due to differences in size of the optical fibers and in lamp housing. Figure \ref{jaipur} shows a comparison between the spectra of the dark Mukundpura meteorite taken with SHADOWS and with the modified version of SHINE.\\

\begin{figure}[h!]
\begin{center}
\includegraphics[scale=0.9]{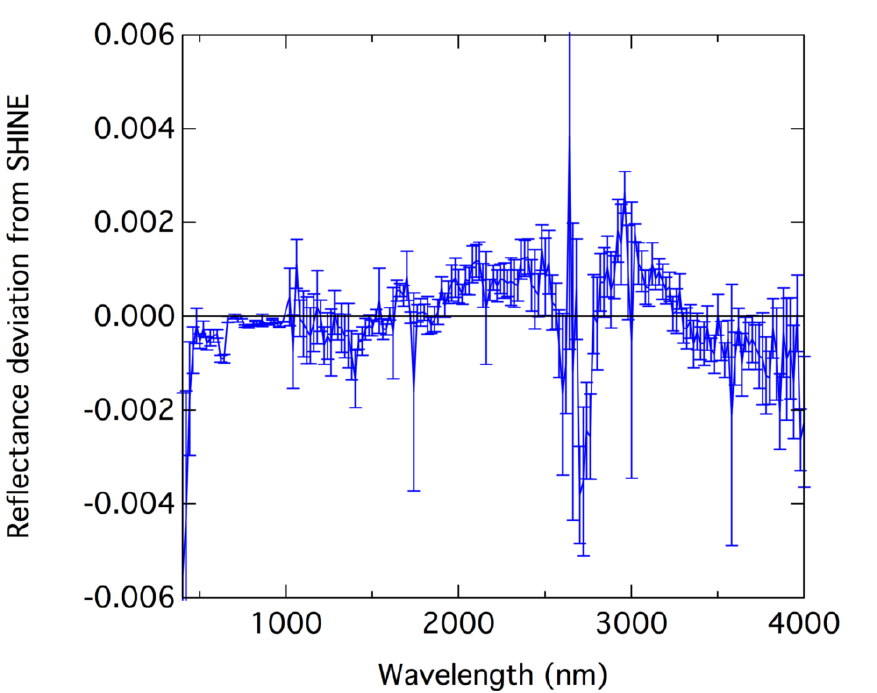}\\
\caption{Differences between the reflectance spectra measured by SHINE and SHADOWS on the Mukundpura meteorite.}
\label{ecart-jaipur}
\end{center}
\end{figure}

\hspace*{0.5cm}Using the same measurement method and exactly the same parameters with both instruments, the reflectance values are very similar for both goniometers, except in the water absorption band at 2.8$\mu$m, better defined in the SHADOWS spectrum. The differences between the reflectance spectra measured with SHADOWS and SHINE are displayed in figure \ref{ecart-jaipur}. The higher difference zone around 2.7$\mu$m corresponds to the spectral range of atmospheric water absorption. Before 500 nm and after 3500 nm, the differences between the spectra are increased by the low signals. Outside these spectral range, the spectrum acquired with SHADOWS shows a deviation from the spectrum measured with SHINE less than \textbf{0.1$\%$}. The spikes and noise are mostly due to SHINE because of its lower SNR.\\

\subsection{Reduced illumination}
\begin{figure}[h!]
\begin{center}
\includegraphics[scale=0.45]{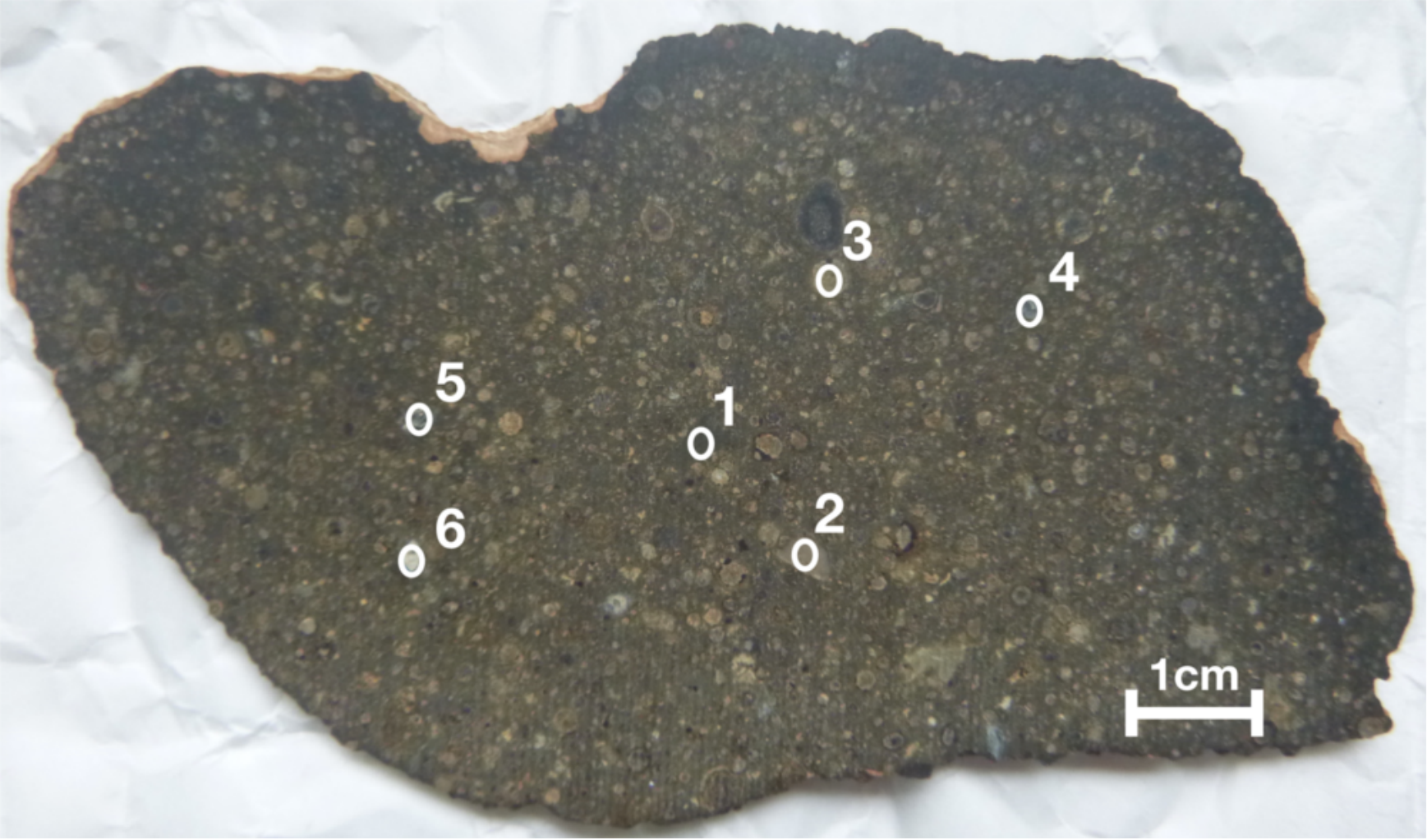}
\caption{Picture of the CV meteorite anoted with the locations of the illumination spot for the six reflectance spectra. The size of the circles represents approximately the size of the illuminated area.}
\label{photo inclusions}
\end{center}
\end{figure}

The 600 $\mu$m pinhole has been installed at the output of the fibers, and carefully aligned to block all the outcoming light except from one fiber, resulting in an illumination spot of 1.7 by 1.3 mm. Using this configuration, reflectance spectra of a heterogenous CV meteorite have been performed. Figure \ref{photo inclusions} shows the studied meteorite and the six illuminated zones for the reflectance spectra.

\begin{figure}[h!]
\begin{center}
\includegraphics[width=0.47\textwidth]{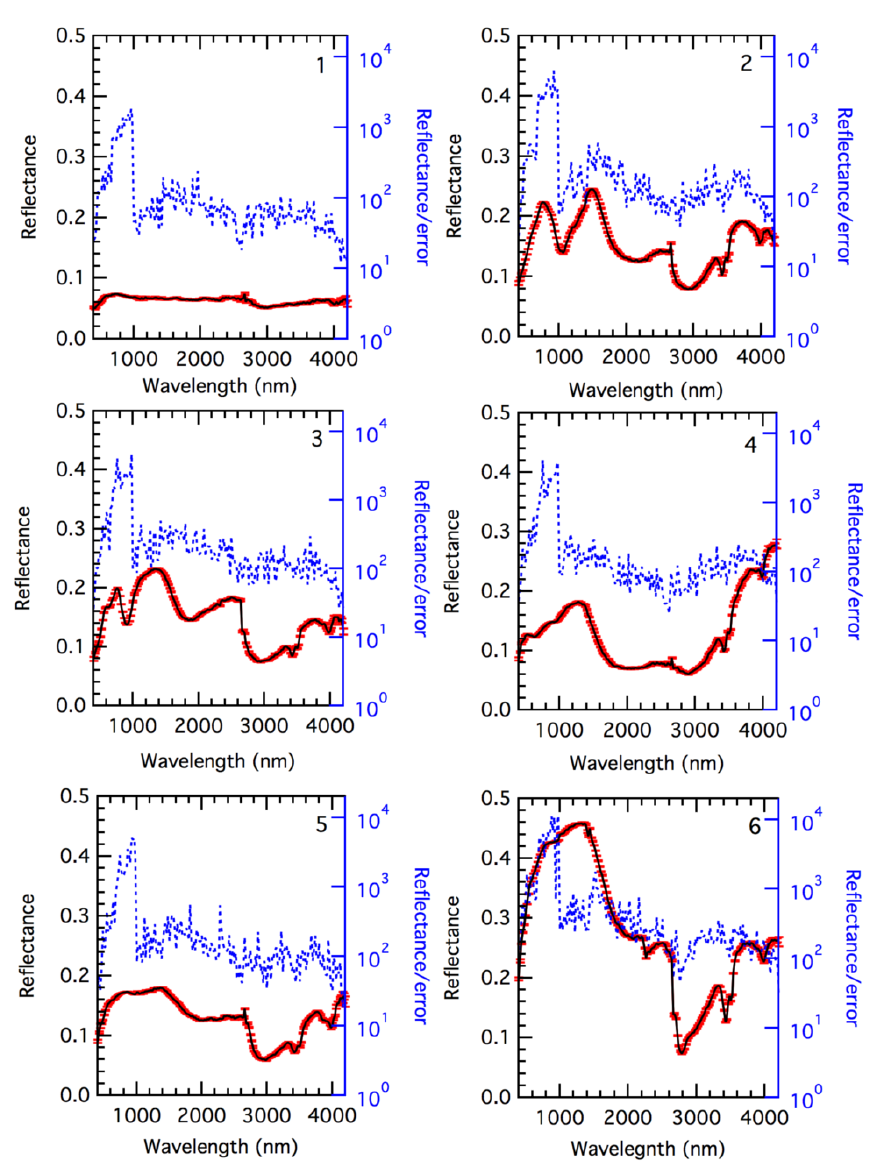}
\caption{Reflectance spectra of different inclusions in the CV meteorite using the 600 $\mu$m pinhole. The spectra are drawn (black) with their error bars (red) and the signal-to-noise ratio is given by the reflectance divided by the associated error (blue).}
\label{spectres_inclusions}
\end{center}
\end{figure}

\hspace*{0.5cm}Figure \ref{spectres_inclusions} show the reflectance spectra of the six inclusions, and the detection signal-to-noise ratio, calculated with the reflectance divided by the associated error, for each measurement. Even for the darkest inclusion, with a mean reflectance around 7$\%$, a signal-to-noise ratio close to 100 over most of the wavelength range is obtained.

\hspace*{0.5cm}These spectra were acquired with only one fiber, but all pinholes can be adjusted to let more than one fiber pass through. This result in a wider illumination spot, but also a higher flux and signal-to-noise ratio.

\subsection{Mechanical adjustments}
\subsubsection{Illumination}
The output of the fibers is installed on an XY translation stage, fixed to the illumination arm with an adaptable stage. The position of the fibers exit can then be adjusted in the three axis. The spherical mirror is held by a kinematic mount that can translate along the arm. Alignment and focusing of the illumination beam was performed using a small CMOS camera whose detector was centered in the center of the sample surface (intersection point of the illumination-observation rotation axis and azimuth rotation axis). The software that came with the camera was used to monitor the illumination spot. The illumination is considered to be Z-aligned when the spot is focused on the CMOS detector and its diameter matches the calculated value. XY alignment was performed by rotating the illumination arm from 0° to 80° and back while checking that the center of the illumination spot did not move on the surface.

\subsubsection{Observation}
Each detector, visible and infrared, can be adjusted to the focal point of their lens assembly by a screw. The adjustment of the detectors at the focal points then consists in optimizing the signal of the detectors. The detectors are considered as set when the measured signal reaches a plateau of maximum intensity, for any angle of observation and azimuth. The adjustment of the XY position of the center of the observation spot is made by 3 screws on each detector support. Alignment is optimized when measurements at an observation angle of 70 ° all return the same intensity value for different azimuth angles.

\subsubsection{Sample holder}
The surface of the sample should be centered at the exact intersection of the rotation axes of the three angles, incidence, emergence and azimuth. Height differences between the surface and the illumination-observation rotation axis can induce photometric errors at high illumination angles, since a poorly focused illumination spot on the sample can partially get out of the observation area. The sample holder is fixed with a screw on the goniometer and can be adjusted vertically with a precision greater than 0.5 mm. A height scale has been especially added to facilitate adjustement which must be done for each sample or reference target. 

\subsection{Transmission mode}
\subsubsection{Performances}
\begin{figure}[h!]
\begin{center}
\includegraphics[scale=1]{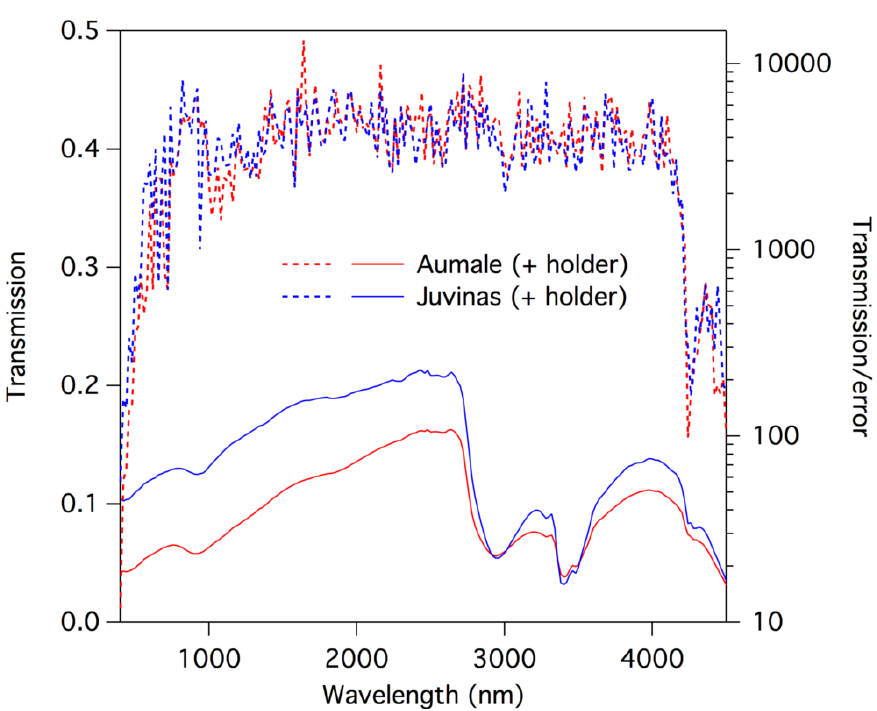}
\caption{Direct transmission spectra of two thin samples of the Juvinas and Aumale eucrite meteorites and associated SNR (ratio between the mean of 10 transmission measurements and 1-sigma standard deviation). The deep absorption features around 2900 nm and 3400 nm in the spectra come from the sample holder. As in reflectance mode, the spectrum is sampled every 20 nm with a spectral resolution varying from 4.8 to 38.8nm. The total measurement time per spectrum is 40 min.}
\label{transmi-lames-minces}
\end{center}
\end{figure}

For diffuse transmission the detectors collect the light scattered through the sample. The intensity of scattered light escaping from the backside of the sample will strongly depend on the absorptivity and scattering efficiency of the material, on the thickness of the sample and on the observation angle relative to the illumination axis. Thus, the dynamic range of these measurements can be quite large, covering several orders of magnitude, in particular for weakly scattering samples with a strong signal in the illumination axis and very weak outside the illumination beam axis. However, given the high sensitivity of SHADOWS, it should be able to detect signal over 4 to 5 orders of magnitude.
In the purely transmissive mode, the light is sent directly to the detectors, passing through the sample. Due to the focus of the illumination beam on the sample the front lens of the detector collects only part (~55$\%$) of the diverging beam (2.9° half angle). To analyse the performances on real samples, two thin slabs of eucrite meteorites, Juvinas and Aumale, were studied in transmission mode (figure \ref{transmi-lames-minces}).

\begin{figure}[h!]
\begin{center}
\includegraphics[width=0.45\textwidth]{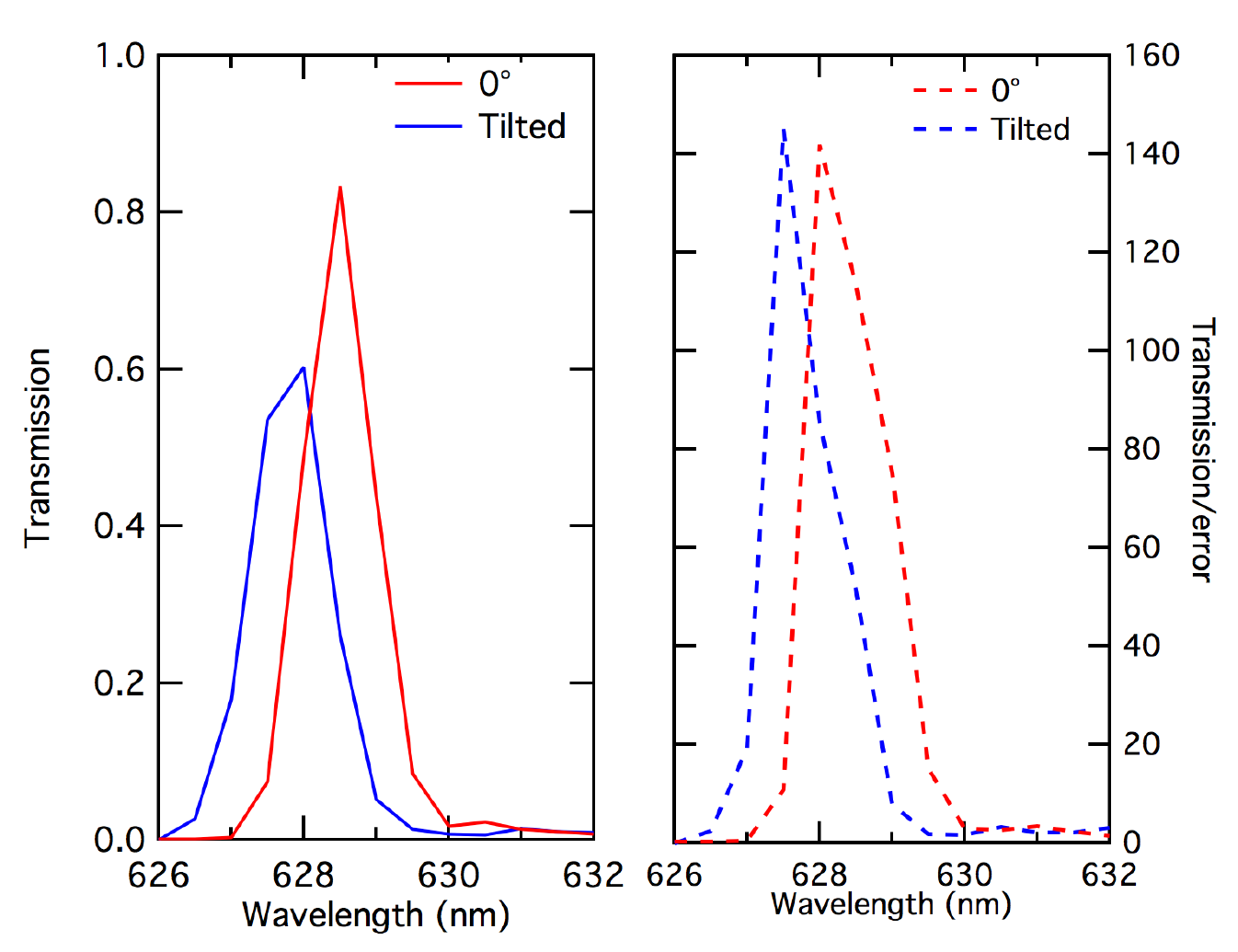}
\caption{Direct transmission spectra and associated SNR of a band-pass filter at normal incidence to the surface and tilted by about 15°. Spectra were acquired with spectral steps of 0.5 nm and a spectral resolution fixed at 0.3 nm. The temperature is -5°C. The SNR corresponds to the ratio between the average of 10 transmission measurements and their 1-sigma standard deviation.}
\label{transmi-filtres}
\end{center}
\end{figure}

\hspace*{0.5cm}The two meteorites show absorption bands around 950nm and 1900nm corresponding to pyroxene features \cite{eucrites}, and absorption bands around 2900nm and 3400nm. These last two come from the resin of the sample holder and are used to test the performances of the instrument in the case of deep absorption bands. With a SNR oscillating aroung 4000, SHADOWS can easily measure transmission spectra over the 350-4500 nm range of highly absorbent samples, with a transmittance of less than a few percents. \\
\hspace*{0.5cm}Due to the high light flux sent directly to the detectors, the transmission mode allows measurements at spectral resolution of 1 nm or even lower. Figure \ref{transmi-filtres} shows the transmission spectra of a band-pass filter, with normal illumination and the filter surface tilted by 15°. 
 
\subsubsection{Limitations}
SHADOWS was not originally designed to measure transmission spectra and is not optimized for this kind of measurement. Due to the horizontal position of the arms and the light path in transmission mode, the sample must be placed vertically in the beam, which reduces the possible measurements to compact samples such slabs, crystals, filters... The transmission of powders cannot be studied with SHADOWS if they are not compacted into a pellet or a thin slab, or contained in a transmission cell. Given the size of the illumination spot the samples must be wider than 6 mm, or 2 mm with the reduced illumination spot. \\
\hspace*{0.5cm}Due to the high direct flux in transmission mode, especially for the 'blank' reference measurement (without sample), it is necessary to reduce the intensity of the monochromatic light to avoid saturation of the detectors. Either the power of the lamp can be reduced, or the monochromator slits can be closed (with the corresponding increase in spectral resolution). This needs to be adjusted in advance by ensuring that there will be no saturation when acquiring the entire spectrum.

\section{Bidirectional reflectance of a challenging surface}
\subsection{Reflectance spectroscopy}
\begin{figure}[h!]
\begin{center}
\includegraphics[scale=0.45]{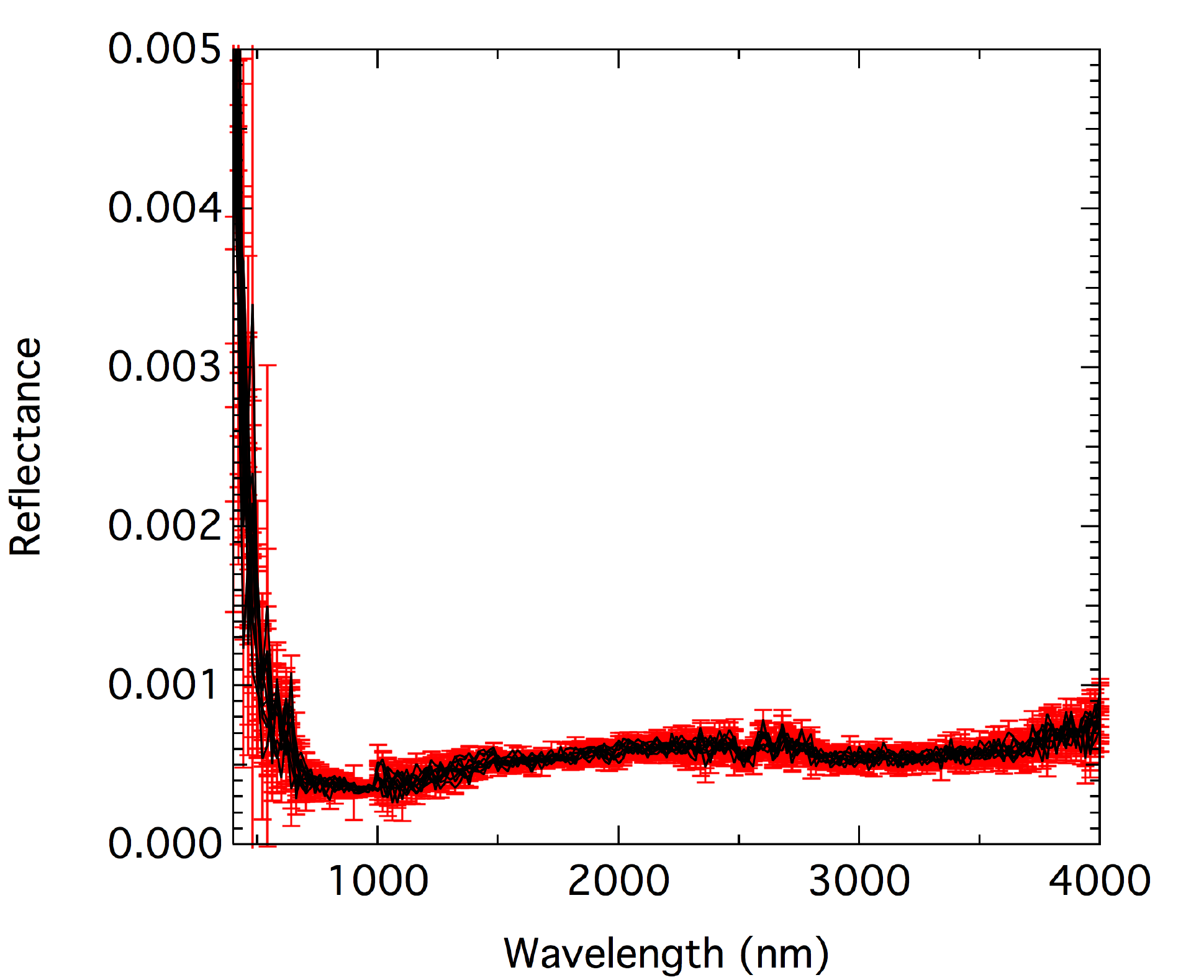}
\caption{Ten reflectance spectra (black) of the specular Vantablack acquired with SHADOWS with a nadir incidence and an observation angle of 30°. The reflectance in the visible is 0.00035. The error bars (red) of each spectra are drawn. Note the reflectance scale between 0 and 0.5$\%$.}
\label{vantablack}
\end{center}
\end{figure}

To push the instrument to its limits, reflectance spectra of the extremely dark surface VANTABlack \cite{vantablack} were tried. The "specular VANTABlack" sample (Surrey NanoSystems VBS1004) available at IPAG is composed of carbon nanotubes grown on aluminum foil . \\
\hspace*{0.5cm}A serie of 10 reflectance spectra was acquired with SHADOWS with a nadir incidence and an emergence angle of 30°. Due to the extreme absorbance of this material, the intensity of the light source has been increased from 150$\mu$A, corresponding to a power of 180W  to 180$\mu$A, so a power of 220W, and the time-constant of the lock-in amplifiers has been changed from 300 ms to 1 s. Spectra are shown in figure \ref{vantablack}. \\

\hspace*{0.5cm}The increase of reflectance in the visible is an artefact due to the very weak signal and the offset generated under these conditions by the instability of the phase measured by the lock-in amplifier, as shown in figure \ref{non linéarité}. The error bars confirm the low signal-to-noise ratio in this part of the spectrum. With this geometry, the reflectance of the VANTABlack is about 0.05$\%$ with a minimum of 0.035$\%$ at 950 nm.
\subsection{Geometrical dependencies}
\begin{figure}[h!]
\includegraphics[width=0.5\textwidth]{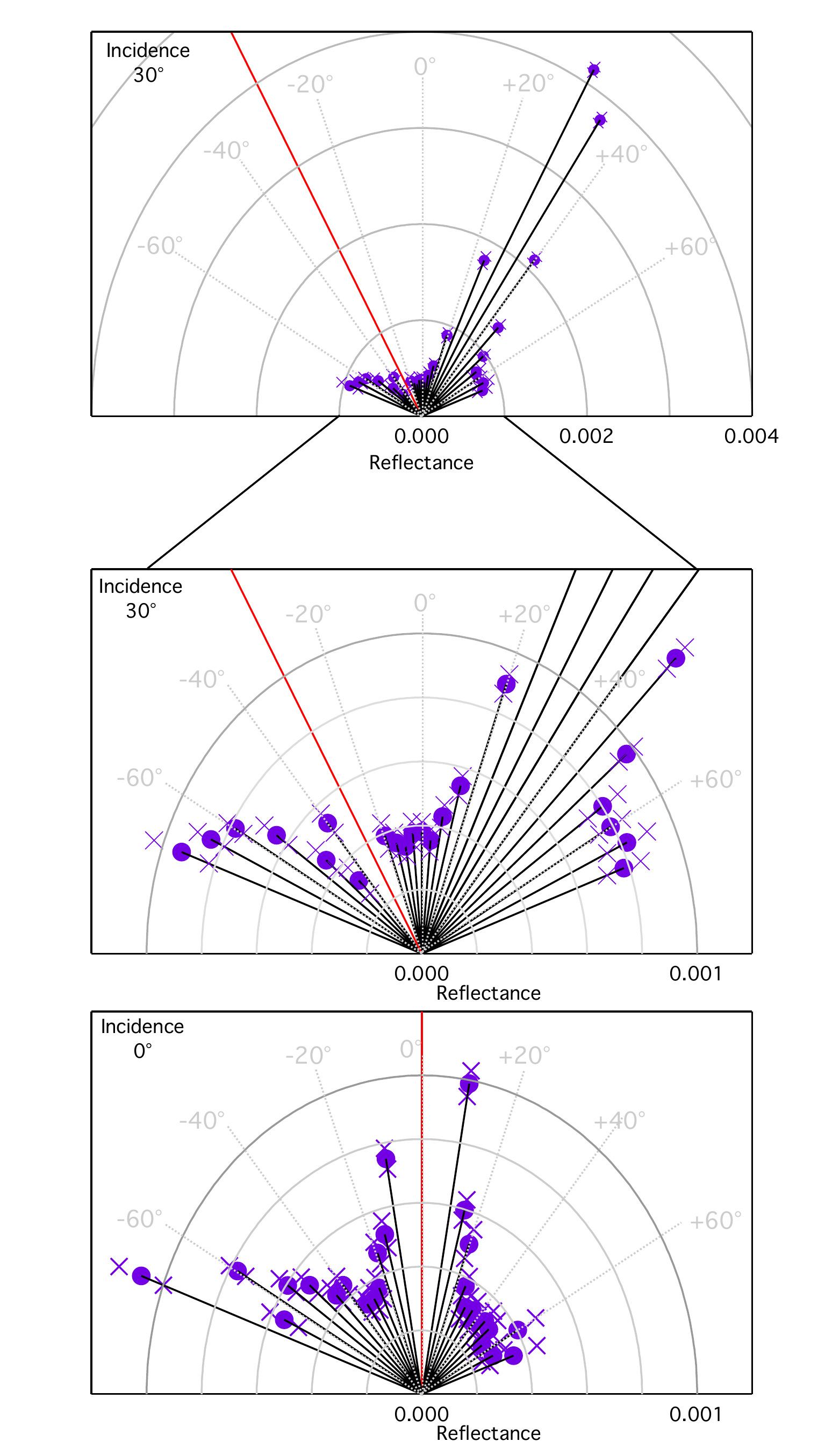}
\caption{BRDF at 1500 nm of the 'specular Vantablack' measured at nadir and 30° incidences. Reflected light is measured in steps of 5° at emergence angles of -70° to 70° with a minimum phase angle of 10° around the illumination. The illumination angle is represented by the red line. The purple dots are the measured reflectance and their $\pm$ 1-sigma error are represented by the two crosses. The gray circles are the reflectance scale of this plot in polar emergence coordinates with the outer circle corresponding to a reflectance of only 0.1$\%$. Reflectance up to 0.4$\%$ occur in the specular peak (values for emergence angles between 25° and 40° are not shown).}
\label{BRDF Vantablack}
\end{figure}

The VANTABlack, and other light-absorbing surfaces, are usually used as stray-light absorbers. The BRDF of such surfaces are thus often needed for optical calculations of the remaining stray-light contribution to the measured signal. We measured the spectral BRDF of the "specular VANTAblack" material at nadir incidence and at 30° incidence angle. The reflected light was measured at emergence angles from -70° to 70° every 5°, except at the angle of illumination and $\pm$ 10° around due to the limitation in phase angle and the absence at that time of a bafle around the output of the fibers limiting the light to be directly sent into the detectors at phase angle 5°. Measurements before 600 nm and after 3500 nm were removed from the nominal spectral range because of the low light intensity in theses ranges. With a time-constant of the lock-in amplifiers of 1 s, the acquisition of the whole set of spectra took roughly 3 days and nights.\\
\hspace*{0.5cm}Figure \ref{BRDF Vantablack} presents the measured BRDF of the VANTABlack at 1500 nm, for two different incidence angles: nadir and 30°.

\begin{figure*}[h!]
\begin{center}
\includegraphics[scale=0.7]{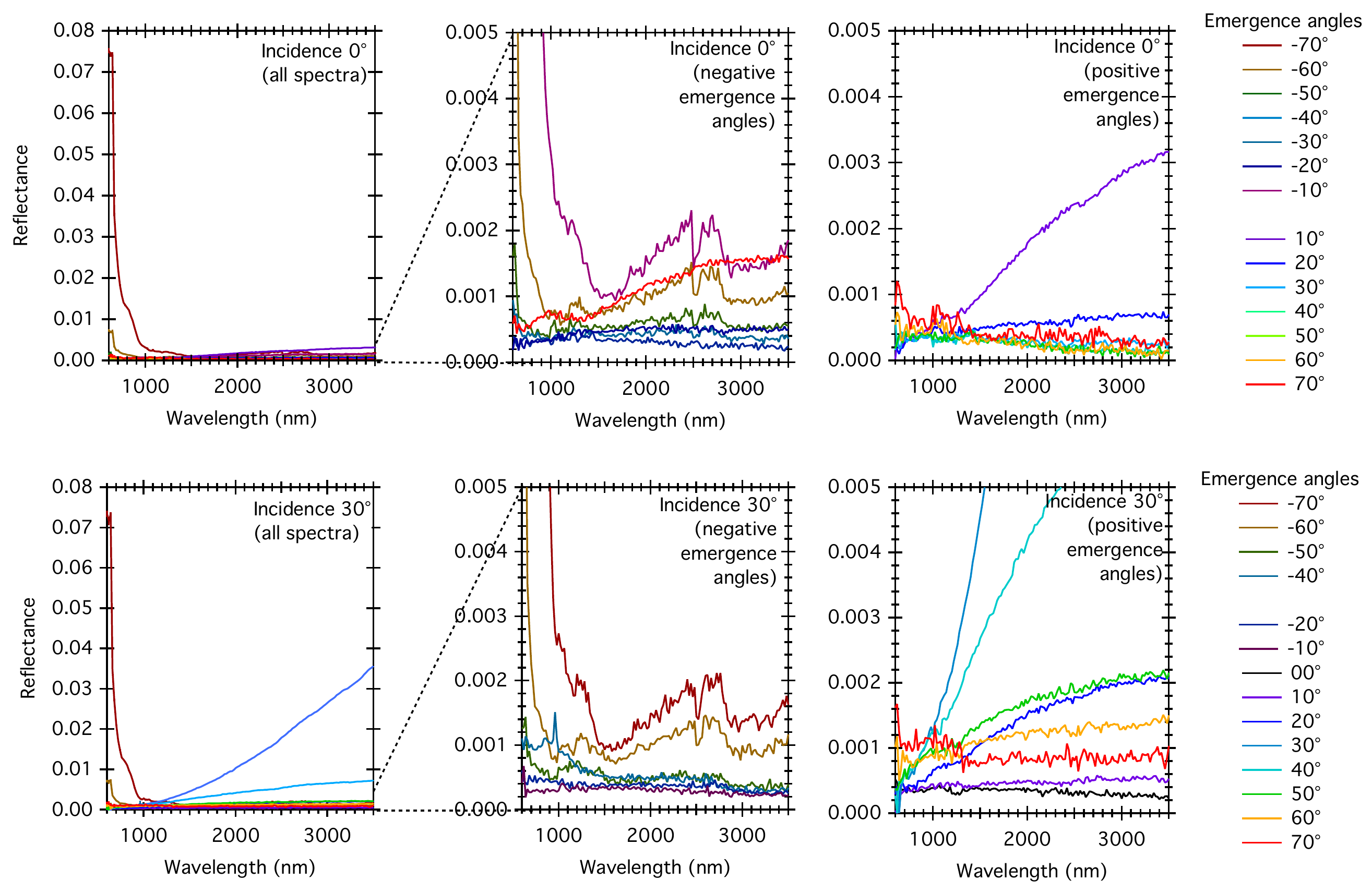}
\caption{Bidirection reflectance spectra of the specular Vantablack at incidence 0° (nadir) and 30°. A selection of spectra are displayed over the whole reflectance range (left), and over a limited range between 0 and 0.5$\%$ to show the low reflectance data in more detail (right).}
\label{spectres_vanta}
\end{center}
\end{figure*}

\begin{figure}[h!]
\begin{center}
\includegraphics[scale=0.4]{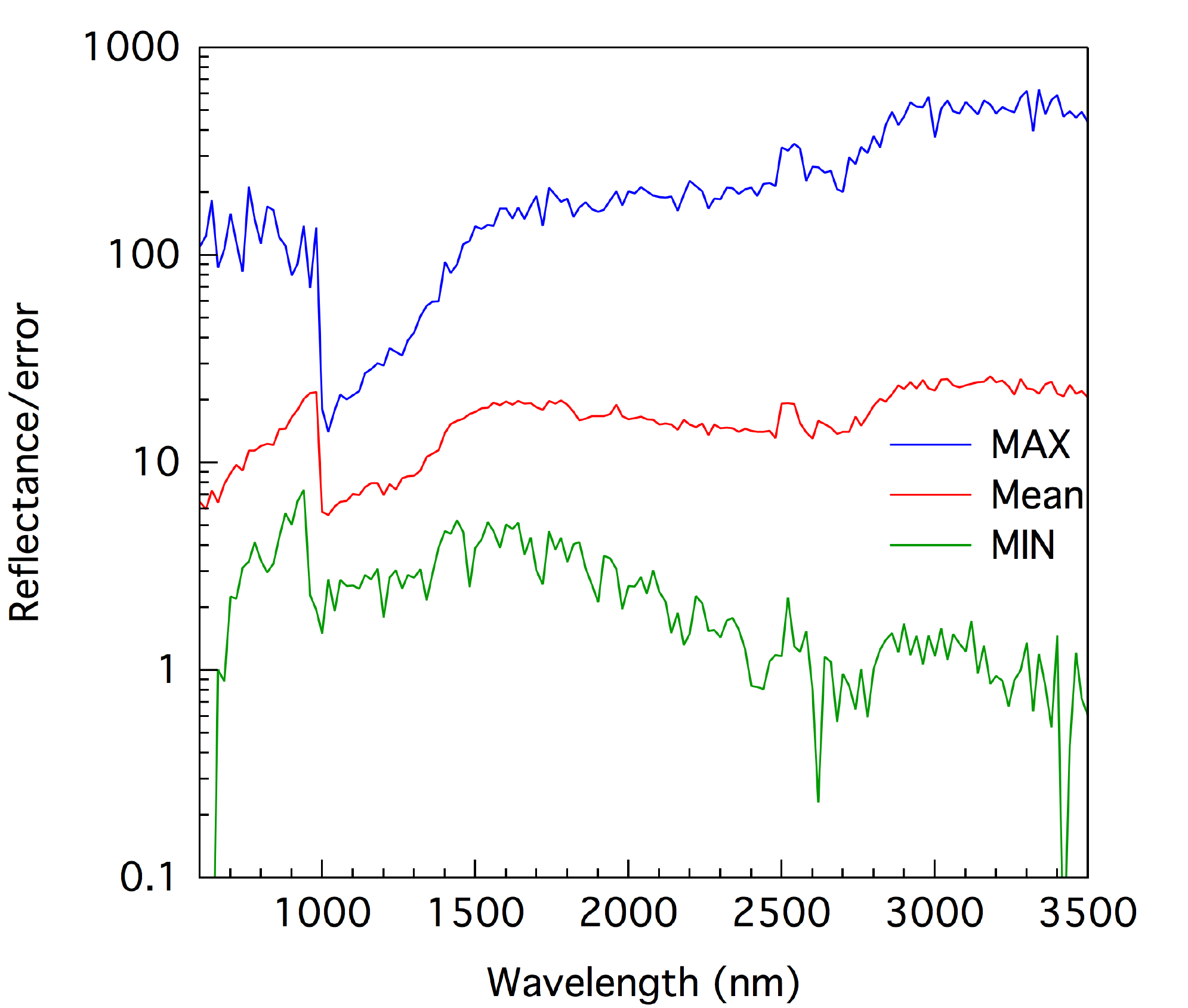}
\caption{Minimum (green), mean (red) and maximum (blue) value of the detection signal-to-noise ratio of all the bidirectional spectra of the VANTABlack.}
\label{SNR Vantablack}
\end{center}
\end{figure}

\hspace*{0.5cm}The specular reflectance peak is up to 0.4$\%$ at the displayed wavelength of 1500nm. The asymetrical shape of the BRDF at nadir illumination may be explained by the substrate being slightly folded or tilted (by 2-3°), dust particles out of the illumination area but scattering the light refelcted by the VANTABlack, or a non-symetrical structure of the surface.\\
\hspace*{0.5cm}The acquired reflectance spectra of the specular VANTABlack at incidences 0° and 30° are displayed on figure \ref{spectres_vanta}. The minimum, maximum and mean values of the detection signal-to-noise ratio for all reflectance spectra of the VANTABlack are represented by figure \ref{SNR Vantablack}. To calculate the mean signal-to-noise ratio for the all 52 reflectance spectra, the minimum and maximum values were removed from the set.\\

\hspace*{0.5cm}BRDF on figure \ref{BRDF Vantablack} show monochromatic photometric variations according to the geometry, but do not display other spectral variations such as a modification of the slope or absorption bands. Study of the whole set of spectra show a drastic variation of the slope between the backward and forward scattering. At grazing observation, the reflectance spectra display an increasing value of reflectance in the visible, higher than 7$\%$. This effect has also be detected at nadir incidence. Around the specular direction, the reflectance is characterized by a steepening of the spectral slope with increasing wavelength, and a reddening while approaching the specular reflection. The reflectance value can go up to 3.6$\%$ in the case of the specular reflection. This reddening effect is also detected at lower level in the case of a nadir illumination. The non-symetrical behaviour detected on the BRDF can be seen on this figure by the similarities between the spectra at grazing observation for both incidence angles.\\
\hspace*{0.5cm}The high signal-to-noise ratio of SHADOWS allows measurement of extremely low reflectance values in a wide range of angular configurations, even near the grazing observation. For the VANTABlack, the highest values of the detection signal-to-noise ratio occur at grazing emergences and for the specular geometry where the reflectance is over 1$\%$, the lowest values of signal-to-noise ratio occur at the darkest parts of the spectra at incidence 0°. The mean value of the detection signal-to-noise ratio, represented by the ratio between the calculated reflectance and the associated error, is around 10, again measuring the reflectance of the darkest surface ever made \cite{darkest_surface}.

\section*{Conclusion}
We presented the design and performances of our new spectro-gonio radiometer capable of measuring the bidirectional reflectance distribution functions of dark and precious samples such as meteorites or terrestrial analogues thanks to its small illumination spot. The two moveable arms allows flexibility over the geometry of the system. The incidence angle can be set to any configuration from a nadir illumination to a 75° angle, and the scattered light can be measured from almost any position on half a hemisphere above the sample thanks to the arm rotating along the emergence and azimuth angles. The instrument is placed in a cold room to acquire spectra from room-temperature down to -20°C. For lower temperature environments, a cryogenic cell is currently under development. The light beam is depolarized to avoid instrumental effects and sample response dependent on polarization. The high signal-to-noise ratio achieved makes it possible to measure the extremely low reflectance levels of dark surfaces, such as the VANTABlack coating. However, limitations are met for millimeter sized samples, as the size of the crystals will be approaching the diameter of the illumination spot, thus reducing the number of illuminated crystals and the measurement will not be statistically relevant. Specifically designed for meteorites and other precious samples, SHADOWS can also be used to perform bidirectional reflectance measurements on artificial dark surfaces, such as VANTABlack or other stray-light absorbing materials. \\
\hspace*{0.5cm}The study of small and precious samples, or small inclusions, is made possible by further reducing the size of the illumination spot while maintaining a good signal-to-noise ratio. Precise photometric measurements can still be performed on just a few ${mm}^{3}$ of dark and fine grained material.\\
\hspace*{0.5cm}This instrument can also be used in transmission mode, be it diffuse or direct, although this is not its original purpose. When the material has a high direct transmittance, measurements with a spectral resolution of less than 1 nm can be obtained, but only with samples that can be installed vertically in the goniometer.\\

\section*{Funding}
This work is part of the Europlanet 2020 RI project which has received funding from the European Union's Horizon 2020 research and innovation programme under grant agreement No 654208. This work was supported by the CNES as a support to several past and future space missions (ESA/Rosetta, NASA/New Horizon, JAXA/Hayabusa, NASA/Mars2020). This work has been supported by a grant from Labex OSUG@2020 (Investissements d’avenir – ANR10 LABX56). Sandra Potin is supported by IRS IDEX /UGA. Pierre Beck acknowledges funding from the European Research Council under the grant SOLARYS ERC-CoG2017$\_$771691.\\ 
\newpage

\bibliography{shadows-arxiv}

\bibliographyfullrefs{sample}
 

\end{document}